\documentclass[12pt,a4paper]{article}
\usepackage[sort&compress]{natbib}
\usepackage[xcolor=svgnames,dvipsnames]{xcolor}
\usepackage{amsmath, amsfonts, amssymb, graphicx}
\usepackage{hyperref}%links table of contents 
\usepackage{dsfont} %math fonts
 \graphicspath{ {./images/} }
\usepackage[margin=1in]{geometry}                
\usepackage{subfig}
\usepackage{adjustbox} %adjust table to fit
\usepackage{cprotect}%protect \verb in another command
\usepackage{verbatim}
\usepackage[autostyle, english = american]{csquotes}
\MakeOuterQuote{"}

\usepackage{algorithm}
\usepackage{algorithmicx}
\usepackage{algpseudocode}

\usepackage{autobreak}
\usepackage{enumitem}% http://ctan.org/pkg/enumitem
\newlist{enumsteps}{enumerate}{2}
\setlist[enumsteps,1]{label={\bf Step \arabic*.~} }
\setlist[enumsteps,2]{label=Step \arabic{enumstepsi}.\arabic* }

\newcommand{\mc}[1]{\mathcal{#1}}

\newcommand{\BM}[1]{\mbox{\boldmath $#1$}}
\newcommand{\mb}[1]{\BM{#1}}

\def\bSig\BM{\Sigma}

\usepackage[figuresright]{rotating}

 \usepackage{authblk}

 \author[1]{Tsering Dolkar\thanks{\href{mailto:tdolkar@vt.edu}{tdolkar@vt.edu}}}
 \author[1]{Marco A. R. Ferreira\thanks{\href{mailto:marf@vt.edu}{marf@vt.edu}}}
 \author[2]{Hwasoo Shin\thanks{\href{mailto:hshin2@hfhs.org}{hshin2@hfhs.org}}}
 \author[3]{Allison N. Tegge\thanks{\href{mailto:ategge@vt.edu}{ategge@vt.edu}}}

 \affil[1]{Department of Statistics, Virginia Tech}
 \affil[2]{Henry Ford Health}
 \affil[3]{Fralin Biomedical Research Institute at VTC, Virginia Tech}

\begin{document}

\title{Bayesian Dynamic Clustering Factor Models}

%\date{{\it Received MM} YYYY. {\it Revised MM} YYYY.  {\it
%Accepted MM} YYYY.}
%\volume{--}
%\pubyear{YYYY}
%\artmonth{MM}
%
%\doi{--------------}
%
%\label{firstpage}
%
\maketitle
\begin{abstract}
We propose novel Bayesian Dynamic Clustering Factor Models (BDCFM) for the analysis of multivariate longitudinal data. BDCFM combines factor models with hidden Markov models to concomitantly perform dimension reduction, clustering, and estimation of the dynamic transitions of subjects through clusters. We develop an efficient Gibbs sampler for exploration of the posterior distribution. An analysis of a simulated dataset shows that our inferential approach works well both at parameter estimation and clustering of subjects. Finally, we illustrate the utility of our BDCFM with an analysis of a dataset on opioid use disorder.
\end{abstract}

\section{Introduction} 
\label{sec:intro}
% last para provides pointers to the sections in the remaining sections. more direct and brief than the previous chapter. 
We propose a novel method for the analysis of multivariate longitudinal data. 
This method is motivated by our previous work on recovery from opioid use disorder (OUD) \citep{craft2023longterm}, where we used a combination of principal component analysis (PCA) and $k$-means clustering~\citep{yeung2001principal,hastie2009elements}. 
While providing dimension reduction and clustering, these latter combined methods do not provide uncertainty quantification. 
We recently developed Bayesian Clustering Factor Models (BCFM), a framework that combines factor models~\citep{LopesWest2004,shin2023dynamic} with mixtures of Gaussians \citep{fruhwirth:2006, fruhwirth2019handbook} to achieve concomitant dimension reduction and clustering for cross-sectional data~\citep{shin2025bayesianclusteringfactormodels}.
To extend this to multivariate longitudinal data, here we propose Bayesian Dynamic Clustering Factor Models (BDCFM).
 
 BDCFM  combines latent factor models (FM) and hidden Markov models (HMM) for concomitant dimension reduction, clustering, and estimation of the dynamic transitions of subjects among clusters. 
 To provide full account of uncertainty, we develop a Markov chain Monte Carlo (MCMC) algorithm~\citep{gelfand:smith:1990, robe:case:2005,game:lope:2006}.
For fast and convenient computations, we consider conditionally conjugate priors. 
To deal with issues of label switching, we employ an empirical Bayes approach inspired by unit information priors~\citep{kass:1995,steele2010performance} to choose the prior hyperparameters for the cluster parameters.

Previous Bayesian approaches have combined FMs with mixtures of Gaussians for concomitant dimension reduction and clustering \citep{fokoue2003mixtures,papastamoulis2018overfitting, chandra2023escaping, ghilotti2025bayesian, shin2025bayesianclusteringfactormodels}. 
These approaches assume for each vector of observations a factor model with vector of common factors coming from a mixture of Gaussian distributions. 
While useful for the analysis of cross-sectional data, these approaches are not adequate for the analysis of longitudinal data. 
To fill this gap, BDCFM assumes that the vector of common factors for each subject evolves through time according to a hidden Markov model. As a result, in addition to providing dimension reduction and clustering, BDCFM also models the dynamic transitions of subjects among clusters. 

We illustrate the usefulness and flexibility of BDCFM with two applications. 
The first application to a simulated dataset shows that our BDCFM framework accurately and precisely estimates the parameters of the model and the cluster assignments. 
The second application, to a longitudinal dataset on recovery from OUD, shows that our BDCFM framework may provide useful information on the latent factors related to the recovery process, the recovery subgroups, and the transitions of subjects among subgroups. 

The remainder of this paper is organized as follows.  
Section \ref{sec:methods} introduces BDCFM and the conditionally conjugate prior distributions. 
Section \ref{sec:fullconditionals} presents the MCMC algorithm for exploration of the posterior distribution, including the full conditional distributions used in a Gibbs sampler. 
Section \ref{sec:simulations} evaluates the performance of our BDCFM framework using a simulated dataset.  
 Section \ref{sec:realdataset} illustrates the application of BDCFM to a real-world dataset on recovery from OUD. 
  Finally, Section \ref{sec:discussion} concludes with a summary of our findings and possible future research directions.
The Appendix contains the empirical Bayes specification of priors for cluster parameters and development of the full conditional distributions.
 Supplementary Table 1 provides the notation used throughout this paper.

\section{The Model}
\label{sec:methods}
\subsection{Bayesian Dynamic Clustering Factor Models}
Consider an $R$-dimensional data vector $\BM{y}_{it}$ for subject $i$ at time $t$, where 
$i = 1,\dots,S$ and $t= 1,\dots,T$. 
Thus, we have $S \times T$ observations for each of the $R$ variables. BDCFM relates the $R$ observed variables to $L$ latent factors using the factor model
\begin{equation}
       \BM{y}_{it} = \BM{B} \BM{x}_{it} + \BM{\epsilon}_{it} \text{, \; } \BM{\epsilon}_{it} \sim N(0, \BM{V}),\\ 
\label{eq1}
\end{equation}
where $\BM{B}$ is an $R \times L$ factor loadings matrix, and $\mb{x}_{it}$ is an $L$-dimensional vector of latent common factors for subject $i$ at time $t$. 
In addition, the covariance matrix of the error term $\mb{\epsilon}_{it}$ is $\BM{V} = diag(\sigma_1^2,\hdots,\sigma_R^2)$, where $\sigma_1^2,\hdots,\sigma_R^2$ are known as idiosyncratic variances or uniquenesses, henceforth referred to as uniquenesses. 
The factor loadings matrix $\mb{B}$ is associated with how much of the variance in an observed variable is explained by the latent common factors. To ensure identifiability of the model, we assume that the factor loadings matrix $\mb{B}$ follows a hierarchical structural constraint such that
$$\mb{B}
= 
\begin{bmatrix}
    1 & 0 & 0 & \dots  & 0 \\
    b_{2,1} & 1 & 0 & \dots  & 0 \\
    b_{3,1} & b_{3,2} & 1 & \dots & 0\\
    \vdots & \vdots & \vdots & \ddots & \vdots \\
    b_{L,1} & b_{L,2} & b_{L,3} & \dots & 1\\
    b_{L+1,1} & b_{L+1,2} & b_{L+1,3} & \dots & b_{L+1,L}\\
    \vdots & \vdots & \vdots & \ddots & \vdots \\
    b_{R,1} & b_{R,2} & b_{R,3} & \dots  & b_{R,L}
\end{bmatrix}.
$$
Thus, we assume that the factor loadings matrix $\mb{B}$ in Equation~(\ref{eq1}) is lower triangular with main diagonal elements set to 1. The order of the variables in $\mb{y}_{it}$ is chosen carefully to preserve the properties of the hierarchical structural constraint \citep{geweke:zhou:1996,aguilar1999bayesian}. \cite{PradoFerreiraWest2021} and \cite{shin2023dynamic} provide recommendations on how to choose the order of the variables. 

Let $Z_{it}$ be a latent variable that indicates the cluster to which subject $i$ belongs at time $t$. 
Denote by $G$ the total number of clusters.
BDCFM assumes that if subject $i$ belongs to cluster $g$ at time $t$, then $Z_{it} = g$ and $\BM{x}_{it}$ follows a multivariate Gaussian distribution with mean vector $\BM{\mu}_g$ and covariance matrix $\BM{\Omega}_g$, that is
$\mb{x}_{it}|Z_{it} = g \sim N(\BM{\mu}_g, \BM{\Omega}_g)$. 
To make BDCFM identifiable,  we assume the first cluster has a diagonal covariance matrix $\BM{\Omega}_1$. 
In addition, BDCFM assumes that $Z_{it}$ follows a hidden Markov model with the transition probability from cluster $j$ at time $t-1$ to cluster $g$ at time $t$ equal to $q_{jg} = P(Z_{it} = g | Z_{i,t-1}= j)$, $t=2,\dots,T$. 
Further, at time $t=1$, the probability of being assigned to cluster $g$ is $p_{g} = P(Z_{i1} = g)$. 
Thus, we define $\BM{Q}$ to be the $G \times G$ matrix of transition probabilities, where the element in row $j$ and column $g$ is $q_{jg}$.

Therefore, the joint probability density function of all the random quantities in BDCFM is 
\begin{eqnarray*}
p(\BM{Y}, \BM{B}, \BM{V}, \BM{X}, \BM{Z}, \BM{\Omega}, \BM{\mu}, \BM{p}, \BM{Q}, \BM{\tau}^2) = \left[\prod_{i=1}^S \prod_{t=1}^T p(\BM{y}_{it} \mid \BM{B}, \BM{x}_{it}, \BM{V})p(\BM{x}_{it} \mid Z_{it}, \BM{\Omega}, \BM{\mu})  p(Z_{it} \mid \BM{p}, \BM{Q})\right]\\
\left[\prod_{g=1}^G p(\BM{\Omega}_g) p(\BM{\mu}_g)  p(\BM{q}_{g.})\right]p(\BM{p}) p(\BM{B} \mid \BM{\tau}^2) p(\BM{\tau}^2),    
\end{eqnarray*}
where $\BM{p} = (p_1,\dots,p_g,\dots,p_G)$, $\BM{q}_{g.} = (q_{g1},\dots,q_{gG})$, and $\BM{\tau}^2 = (\tau^2_1,\dots,\tau^2_L)$. In addition, $p(\BM{\Omega}_g)$, $p(\BM{\mu}_g)$, $p(\BM{q}_{g.})$, $p(\BM{p})$, $p(\BM{B}\mid \BM{\tau}^2)$, and $p(\BM{\tau}^2)$ are prior densities; we discuss the choice of prior distributions in the next section.

\subsection{Prior distributions}

We propose using conditionally conjugate prior distributions that allow a wide range of prior information to be represented and facilitate MCMC computations. 

For the factor loadings matrix $\mb{B}$, we assign independent normal priors with mean 0 and variance~$\tau_l$ to the unconstrained elements of the $l$-th column (the $l$-th factor).
In addition, we assign for the variance $\tau_l$ an inverse gamma prior distribution $IG(n_{\tau}/2, n_{\tau}s_{\tau}^2/2)$, where $n_\tau = 1$, and $n_\tau s_{\tau}^2 = 1$. Further, as recommended by \citet{LopesWest2004}, for the uniqueness variances $\sigma_1^2, \dots, \sigma_R^2$, we assign independent inverse gamma priors $IG(n_\sigma/2,n_\sigma s_\sigma^2/2)$, where $n_\sigma = 2.2$, and $n_\sigma s_{\sigma}^2 = 0.1$. This choice of hyperparameters results in a prior mean of 0.5 and infinite prior variance. 
% \mu_j^* -> m_j^\mu
% \Omega_j^* ->  C_j^\mu

Next, we discuss the prior distributions for the cluster parameters. We assume for the mean vector $\BM{\mu}_g$ of cluster $g$ a normal prior distribution with $\BM{\mu}_g \sim N(\BM{m}_g^\mu, \BM{C}_g^\mu)$, $g=1,\dots,G$. We discuss the choice of $\BM{m}_g^\mu$ and $\BM{C}_g^\mu$ in the Appendix. Recall that the covariance matrix of the first cluster $\BM{\Omega}_1$ is diagonal. Denote by $\omega_l$ the $l$-th diagonal element of $\BM{\Omega}_1$ which corresponds to the $l$-th factor. 
We assume for $\omega_{1},\dots,\omega_{L}$ independent inverse gamma priors with $\omega_{l} \sim IG(n_{\omega_l}/2,n_{\omega_l} s_{\omega_l}^2/2)$. 
We set the shape parameter of the prior to $n_{\omega_l}/2 = 2$, which implies that the prior mean of $l$-th element $\omega_{l}$ is $s_{\omega_{l}}^2$. 
The Appendix discusses the choice of $ s_{\omega_l}^2$. 
Further, we assume that the covariance matrix $\BM{\Omega}_g$ of cluster $g$ follows an inverse Wishart prior with $\BM{\Omega}_g \sim IW(n_{\Omega_g}, \BM{S}_{\Omega_g})$, 
 $g = 2,\dots, G$ where $n_{\Omega_g} = 2 + L$. 
 This choice of hyperparameters implies that the prior mean of $\BM{\Omega}_g$ is  $\BM{S}_{\Omega_g}$.
We discuss the choice of $\BM{S}_{\Omega_g}$ in the Appendix.
 
 Lastly, we discuss the prior distributions for the initial cluster probabilities and the transition probability matrix. 
 Recall that the probability of a subject starting in cluster $g$ at time point 1 is $p_g$. 
 We denote the vector of initial probabilities by $\mb{p} = (p_1,\dots,p_G)$. We assume a \textit{priori} that $\mb{p}$ follows a Dirichlet distribution with parameter vector $(\alpha_1,..,\alpha_G) = (2,...,2)$. 
 Let $q_{j.} = (q_{j1},\dots,q_{jG})$ be the vector of probabilities of transitioning from cluster $j$ at one-time point to each of the $G$ clusters at the next time point. 
 Note that $q_{jj}$ is the probability of staying in the cluster $j$ at the next time point and that  $q_{j1} + \dots + q_{jG} = 1$. We assume that $q_{j.} \sim Dirichlet(\alpha_{j1},\dots,\alpha_{jG})$ where $\alpha_{jg}=2$, $g=1,...,G$. 
 Note that both for the initial probabilities and the transition probabilities, the choice of exponents $\alpha = 2$ implies that the prior density is equal to zero if any element of the probability vector is equal to zero. 
 
 % We have chosen these values based on a simulation study with the Dirichlet distribution. Specifically, these values for $(\alpha_{l1},\dots,\alpha_{lG})$ imply that the prior mean of the probability of staying in the same cluster at the next time point is greater than 30\%. We chose a function of $G$ as prior for the probability of staying in the same cluster because as $G$ increases, $\alpha_{ll}$ increases as well, which increases the concentration parameter for the current cluster assignment relative to the others. This prior choice favors the existence of stable clusters.

\section{Statistical Inference}
\label{sec:fullconditionals}
We propose a Gibbs sampler \citep{gelfand1990sampling} to obtain a sample from the joint posterior distribution of all the parameters in BDCFM. In this section, we present the full conditional distributions of the parameters and the Gibbs sampler. 

The full conditional distribution of the vector of common factors $\BM{x}_{it}$ depends on the cluster assignment $Z_{it}$ of subject $i$ at timepoint $t$. Specifically, given $Z_{it} = g$, the full conditional of $\BM{x}_{it}$ is $N(\BM{m}_{it}, \BM{C}_{it})$ with
\begin{eqnarray}
    \label{eq8} 
    \BM{C}_{it} & = &(\BM{B}^{'}\BM{V}^{-1}\BM{B}+\BM{\Omega}_g^{-1})^{-1},\\ \nonumber
    \BM{m}_{it} & = & \BM{C}_{it}(\BM{B}^{'}\BM{V}^{-1}\BM{y}_{it} + \BM{\Omega}_g^{-1}\BM{\mu}_g),
\end{eqnarray}
where $\BM{\mu}_g$ and $\BM{\Omega}_g$ are the prior mean vector and the prior covariance matrix of the $g$-th cluster. In addition, recall that $\BM{V} = diag(\sigma_1^2,\dots,\sigma_R^2)$. 

The full conditional distribution of the mean of the $g$-th cluster, $\BM{\mu}_g$, is $N(\BM{m}_g^{\mu*}, \BM{C}_g^{\mu*})$, with 
\begin{eqnarray}\label{eq11}
    \BM{C}_g^{\mu*} &=&(\BM{C}_g^{\mu-1}+ \BM{\Omega}_g^{-1}\sum_{t=1}^Tn_{tg})^{-1},\\ \nonumber
    \BM{m}_g^{\mu*} &=& \BM{C}_g^{\mu*} (\BM{C}_g^{\mu-1} \BM{m}_g^\mu+ \BM{\Omega}_g^{-1} \sum_{t=1}^T \sum_{i \in \mc{C}_{tg}} \BM{x}_{it}),
\end{eqnarray}
where $\mc{C}_{tg}$ is the set of subjects that belong to cluster $g$ at time $t$, and $n_{tg}$ is the cardinality of~$\mc{C}_{tg}$.

Recall that the covariance matrix of the first cluster is $\BM{\Omega}_1 = diag(\omega_1, \dots, \omega_L)$. 
The full conditional distribution of $\omega_{l}$, $l=1,\dots,L$, is $IG(n_{\omega_l}^*/2, n_{\omega_l}^*s_{\omega_l}^{*2}/2)$, with
\begin{eqnarray}
\label{eq9}
n_{\omega_l}^* &=& n_{\omega_l}+\sum_{t=1}^Tn_{t1} , \\ \nonumber
n_{\omega_l}^*s_{\omega_l}^{*2} &=& n_{\omega_l} s_{\omega_l} ^2 + \sum_{t=1}^T\sum_{i \in \mc{C}_{t1}}\left(x_{itl}-\mu_{1l}\right)^2,
\end{eqnarray}
where $x_{itl}$ is the $l$-th element of $\BM{x}_{it}$, and $\mu_{1l}$ is the $l$-th element of the mean vector of cluster~1.

The full conditional distribution of the covariance matrix $\BM{\Omega}_g$ for $g=2, \dots, G$ is an inverse Wishart distribution $IW(n_{\Omega_g}^*, \BM{S}_{\Omega_g}^* )$ with parameters
\begin{eqnarray}\label{eq10}
    n_{\Omega_g}^* &=& n_{\Omega_g} + \sum_{t=1}^T n_{tg}, \\ \nonumber
    \BM{S}_{\Omega_g}^* &=&  \BM{S}_{\Omega_g} + \sum_{t=1}^T \sum_{i \in \mc{C}_{tg}}(\BM{x}_{it}-\BM{\mu}_g)(\BM{x}_{it}-\BM{\mu}_g)'.
\end{eqnarray}

We simulate the factor loadings matrix $\mb{B}$ by row. 
Let $\BM{b}_{r.}$ be the $r$-th row of  $\BM{B}$.
Due to the hierarchical structural constraint, the first row of $\mb{B}$ is fixed.
For the remaining rows of $\mb{B}$, we  consider two cases; Case 1: $1 < r \leq L$, and Case 2:  $r > L$. 
For Case 1,  the last $L - r + 1$ elements of $\mb{b}_{r.}$ are fixed. 
This implies that there are $r-1$ free elements in $\mb{b}_{r.}$. 
Let $\mb{b}_{r.}^*$ be a vector that contains the first $r-1$ elements of $\mb{b}_{r.}$. 
In addition, let $\BM{\Upsilon}=diag(\tau_1,\ldots,\tau_L)$. 
Then for Case 1, the full conditional distribution of $\mb{b}^*_{r.}$ is $N(\mb{m}^b_r, \mb{C}^b_r)$, with 
\begin{eqnarray}
\label{eq13}
    \mb{C}^b_r & = & \left(\BM{\Upsilon}_{1:(r-1),1:(r-1)}^{-1} + \frac{1}{\sigma_r^{2}}\sum_{t=1}^T \sum_{i=1}^S\BM{x}_{it, 1:(r-1)} \BM{x}_{it, 1:(r-1)}^{\prime}\right)^{-1},\\
    \nonumber
    \mb{m}^b_r &=& \mb{C}_r^b \frac{1}{\sigma_r^2} \sum_{t=1}^T \sum_{i=1}^S\BM{x}_{it, 1:(r-1)}(y_{itr}-x_{itr}),
\end{eqnarray}
where $\BM{\Upsilon}_{1:(r-1),1:(r-1)}$ is a submatrix of the first $r-1$ rows and columns of the covariance matrix $\BM{\Upsilon}$. 
For Case 2, $r > L$, the full conditional distribution of  $\mb{b}_{r.}$ is $N(\BM{m}^b_r, \BM{C}^b_r)$, with
\begin{eqnarray}\label{eq12}
    \BM{C}^b_r &=& \left(\BM{\Upsilon}^{-1}+\frac{1}{\sigma_r^2}\sum_{t=1}^T \sum_{i=1}^S \mb{x}_{it} \mb{x}_{it}^{\prime}\right)^{-1},\\ \nonumber
    \BM{m}^b_r &=& \mb{C}_r^b \frac{1}{\sigma_r^2} \sum_{t=1}^T \sum_{i=1}^S \mb{x}_{it} y_{itr}.
\end{eqnarray}

The full conditional distribution of the $r$-th uniqueness $\sigma_r^2$, for $r=1,\dots,R$, is an inverse gamma distribution $IG(n_\sigma^*/2, n_\sigma^*s_\sigma^{*2}/2)$ with
\begin{eqnarray}\label{eq14}
    n_\sigma^* & = & n_\sigma + ST , \\ \nonumber
    n_\sigma^*s_\sigma^{*2} & = & n_\sigma s_\sigma^2 + \sum_{t=1}^T\sum_{i=1}^S(y_{itr}-\BM{x}_{it}^\prime\BM{b}_{r.})^2.
\end{eqnarray}

Now let us consider the full conditional distribution of the variance $\tau_l$ of the factor loadings of the $l$-th factor, $l=1,\dots, L$.
 Due to the hierarchical structural constraint of $\mb{B}$, the first $l$ elements of the $l$-th factor are fixed at either 0 or 1. 
 Thus,  the full conditional distribution of  $\tau_l$ is an inverse gamma distribution $  IG(n_\tau^*/2, n_\tau^*s_\tau^{*2}/2)$, with
\begin{eqnarray}\label{eq15}
    n_\tau^* & = & n_\tau+ R - l, \\ \nonumber
    n_\tau^*s_\tau^{*2} & = & n_\tau s_\tau^2 + \BM{B}_{(l+1):R,l}^{\prime} \BM{B}_{(l+1):R,l},
\end{eqnarray}
where $\BM{B}_{(l+1):R,l}$ is the vector with the last $R-l$ elements of the 
$l$-th column of $\BM{B}$.

Recall that the initial probability that subject $i$ belongs to cluster $g$ at time $t=1$ is $p_g = P(Z_{i1} = g)$. The full conditional distribution of the vector of initial probabilities $\BM{p}=(p_1,\dots,p_G)$ is a Dirichlet distribution,
\begin{eqnarray}\label{eq16}
    \BM{p} \mid. \sim Dir(\BM{n}_{1}+\BM{\alpha}),
\end{eqnarray}
where $\BM{n}_{1}=(n_{11},\dots,n_{1G})$, $n_{1g}$ is the number of subjects in cluster $g$ at time $t=1$, and $\BM{\alpha} = (\alpha_1,\dots,\alpha_G)$. 

Let us now consider the simulation of the transition probability matrix $\BM{Q}$. The full conditional distribution of the $j$-th row of  $\BM{Q}$ is a Dirichlet distribution 
\begin{eqnarray}\label{eq17}
  \BM{q}_{j.} \mid . \sim Dir(\BM{m}_{j.}+\BM{\alpha}_{j.}),
\end{eqnarray}
where $\BM{m}_{j.} = (m_{j1},\dots,m_{jG})$ and $m_{jg} = \sum_{i=1}^S \sum_{t=2}^T \mathds{1}(Z_{it}=g \mbox{ and } Z_{i,t-1}=j)$, and $\BM{\alpha}_{j.} = (\alpha_{j1},\dots,\alpha_{jG})$. 

The full conditional distribution of $Z_{it}$, the cluster assignment of the $i$-th subject at time~$t$, given that $Z_{i,t-1} = j$ and $Z_{i,t+1} = p$, $t=2, \dots, T-1$, is
\begin{eqnarray}\label{eq18}
    P(Z_{it} = g \mid Z_{i,t-1} = j,Z_{i,t+1} = p, .) \propto N(\BM{x}_{it} \mid \BM{\mu}_g, \BM{\Omega}_g) q_{jg}  q_{gp}, 
\end{eqnarray}
where $j=1,\dots,G, g=1,\dots,G$ and $p=1,\dots,G$.
When $t=1$, $    P(Z_{it} = g \mid Z_{i,t+1} = p, .) \propto N(\BM{x}_{it} \mid \BM{\mu}_g, \BM{\Omega}_g)  q_{gp}$. 
When $t=T$, $P(Z_{it} = g \mid Z_{i,t-1} = j, .) \propto N(\BM{x}_{it} \mid \BM{\mu}_g, \BM{\Omega}_g) q_{jg}$.

Using the above full conditional distributions, we simulate from the posterior distribution of the BDCFM parameters using the following Gibbs sampler:

\begin{enumsteps}[leftmargin=.75cm,labelwidth=\itemindent,labelsep=.1cm,align=left]
 
  \item Set initial values of $\BM{x}_{it}, \BM{B}, \sigma_1^2,\hdots,\sigma_R^2,$ $\tau_1,\hdots,\tau_L$, $\BM{\mu}_g$,  $\BM{\Omega}_g$, $\BM{Z}$, and $\BM{Q}$.
  
  \item Simulate $\mb{x}_{it}$, $i=1,\dots,S$, $ t=1,\dots,T$, from its Gaussian full conditional distribution given in Equation~(\ref{eq8}).
  
  \item Simulate $\mb{\mu}_g$, $g=1,\dots,G$, from its Gaussian full conditional distribution given in Equation~(\ref{eq11}).
  
  \item Simulate each $\omega_{1},\hdots,\omega_{L}$ from its inverse gamma full conditional given in Equation~(\ref{eq9}) and $\Omega_2, \hdots, \Omega_G$ from its inverse Wishart full conditional distribution given in Equation~(\ref{eq10}).
  
  \item Simulate the matrix of factor loadings $\mb{B}$ from the Gaussian full conditional distribution given in Equation~(\ref{eq13}) and Equation~(\ref{eq12}).
  
  \item Simulate each $\sigma_1^2,\hdots,\sigma_R^2$ from the inverse gamma full conditional distribution given in Equation~(\ref{eq14}).
  
  \item Simulate each $\tau_1,\hdots,\tau_L$ from the inverse gamma full conditional distribution given in Equation~(\ref{eq15}).

  \item Simulate the vector of initial probabilities $\mb{p}$ from the Dirichlet distribution given in Equation~(\ref{eq16}).
  
  \item Simulate the vector of cluster transition probabilities $\mb{q}_{j.}$ from its Dirichlet full conditional distribution given in Equation~(\ref{eq17}).
  
  \item Simulate $Z_{it}$, $i=1,\dots,S$, $t=1,\dots,T$, from its full conditional distribution given in Equation~(\ref{eq18}).
  
  \item Repeat {\bf Step 2} to {\bf Step 10} until MCMC converges and we have enough posterior samples.
\end{enumsteps}

\section{Analysis of a Simulated Dataset}
\label{sec:simulations}
We have simulated a dataset with $S=200$ subjects, $R=20$ variables, $T=5$ time points, $G=4$ clusters, and $L=3$ factors. The $200$ subjects at the first time point are randomly assigned to one of the $4$ clusters according to the probability vector $\BM{p} = (0.45, 0.26, 0.16, 0.13)$. Then, in the subsequent time points, subjects transition among clusters according to the transition probability  matrix 
% \begin{table}[H]
%     \centering
%     \begin{tabular}{|c|c|c|c|c|c|}
%     \hline 
%      from $\mb{G}$/to $\mb{G}$  & 1 & 2 & 3 & 4\\
%     \hline
%       1 & 0.75 & 0.15 & 0.07 & 0.03\\
%     \hline
%       2 & 0.23 & 0.57 & 0.09 & 0.11\\
%     \hline
%       3 & 0.26 & 0.13 & 0.47 & 0.14\\
%     \hline
%       4 & 0.09 & 0.19 & 0.20 & 0.52\\
%     \hline
%     \end{tabular}
%         \caption{Transition matrix, $\mb{Q}$}
% \end{table} 
% make this a matrix
\[\BM{Q}
= 
\begin{bmatrix}
    0.75 & 0.15 & 0.10 & 0.00 \\
    0.20 & 0.55 & 0.15 & 0.10 \\
    0.25 & 0.15 & 0.50 & 0.10 \\
    0.10 & 0.15 & 0.20 & 0.55
\end{bmatrix}.\]
The true mean vectors of each cluster are $\BM{\mu}_1 = (7, 4, 5)$, $\BM{\mu}_2 = (-7, 3, -3)$, $\BM{\mu}_3 = (6, -3, -2)$, and $\BM{\mu}_4 = (-6, -4, 3)$. The true cluster covariance matrices are
\[ \BM{\Omega}_1
= 
\begin{bmatrix}
    1.9 & 0.0 & 0.0 \\
    0.0 & 1.1  & 0.0 \\
    0.0 & 0.0 & 1.3
\end{bmatrix}
, \BM{\Omega}_2
= 
\begin{bmatrix}
    2.0 & 0.4 & 0.4 \\
    0.4 & 2.0   & 0.4 \\
    0.4 & 0.4 & 2.0
\end{bmatrix},\\
\] 
\[\BM{\Omega}_3
= 
\begin{bmatrix}
    3.0 & 0.6 & 0.6 \\
    0.6 & 3.0 & 0.6 \\
    0.6 & 0.6 & 3.0
\end{bmatrix}\\, \text{ and }
\BM{\Omega}_4
= 
\begin{bmatrix}
    4.0 & 1.0 & 1.0 \\
    1.0 & 4.0 & 1.0 \\
    1.0 & 1.0 & 4.0
\end{bmatrix}.\\
\] 

We have performed a BDCFM analysis of this simulated dataset using the Gibbs sampler described in Section~\ref{sec:fullconditionals}. Specifically, we have run the Gibbs sampler for 50,000 iterations, thinning every 10 iterations and considering a burn-in of 10,000 iterations.

% matrix of true B; remaining figures in appendix
% \begin{figure}[H]
%     \centerline{\includegraphics[width=\linewidth]{proj1/sim-data/B_CI_g4l3s200times5.pdf}}
%     \caption{Simulated data -- factor loadings for the 4-cluster-3-factor dataset with 20 original variables using BDCFM: true value (blue triangle), posterior mean (black dot), and 95\% credible interval (black vertical line).}
%     \label{fig:simulatedB}
% \end{figure}

For this simulated dataset, the inferential procedure for BDCFM accurately estimated the factor loading matrix $\mb{B}$ (Supplementary Figure 1),  the posterior densities of the cluster means $\mb{\mu}_g$, $g=1,\dots, 4$ (Supplementary Figure 2),  
 the cluster transition probabilities $q_{jg}$, $j=1,\dots, 4$ and $g=1,\dots, 4$ (Supplementary Figure 3),
 the initial cluster probabilities $\mb{p}$ (Supplementary Figure 4), and 
 the uniquenesses $\sigma_{r}^2$ for $r=1,\dots,R$ (Supplementary Figure 5). 
%The clusters label on the x-axis of Supplementary Figure 3b has two digits. This figure shows the probability of transitioning from the cluster of the first digit to the cluster of the second digit. As can be seen, all but two of the true values are within the 95\% credible interval and sometimes visually overlap with the posterior mean.
When considering the factor loadings matrix, the cluster means, the transition probabilities, the initial cluster probabilities, and the uniquenesses together,  1.9\% of the 95\% credible intervals did not include the true value. Therefore, the inferential procedure for BDCFM  adequately quantifies the uncertainty. 

% \begin{figure}[H]
%     \centering
%     \includegraphics[width=\linewidth]{proj1/sim-data/subject_time_alluvial_g4s200times5.pdf}
%     \caption{Simulated data -- alluvial plot of cluster assignment for 200 subjects over 5 time-points: BDCFM cluster assignment (left panel) and true cluster assignment (right panel).}
%     \label{fig:simulatedCA}
% \end{figure}

With respect to cluster assignments, BDCFM classifies the subjects almost perfectly with a misclassification rate of 0.1\%. In addition, 
Supplementary Figure 6 shows the alluvial plot of the cluster assignments for 200 subjects over 5 time points as predicted by BDCFM compared to the true assignments. 

\section{BDCFM analysis of OUD recovery}
\label{sec:realdataset}
We analyze data from a  24-month observational study of patients  who had received extended-release buprenorphine for the treatment of OUD (NCT03604861). 
We perform a BDCFM analysis of a cohort of $S=252$ participants who completed assessments at $T=4$ time points (6, 12, 18, and 24 months). Due to protocol changes, time point 0 did not have complete data and was thus excluded from this analysis. 
We consider 15 variables: three variables from the Sheehan Disability Scale (SDS) related to impairment in work, in social life, and in family life \citep{sheehan1998mini}; Kessler Psychological Distress Scale (K6) \citep{kessler2003screening}; Subjective Opioid Withdrawal Scale (SOWS) \citep{handelsman1987two}; three variables from the Brief Pain Inventory (BPI) that provides information regarding pain on average, pain at its worst, pain at its best \citep{mendoza2006reliability}; Beck Depression Inventory-II (BDI-II) \citep{beck1996bdi}; two variables from Short Form Health Survey (SF-12), which considers a 12-item measure of health-related quality of life, that here is separated in SF12 mental score and SF12 physical score \citep{ware1996sf12}; one variable regarding the number of days in conflict in family and social relationships; one variable about number of medicines or drugs used for pain in the past 30 days; one variable regarding patient satisfaction towards treatment; and one variable regarding patient confidence about remaining abstinent from abusing opioids. 
Here, we perform a BDCFM analysis with $G=4$ clusters and $L=3$ factors as  identified by our prior study~\citep{craft2023longterm}. The BDCFM analysis took 6.2 minutes to run.

% \begin{table}[ht]
% \centering
% \caption{Factor Loadings for Latent Variables (L1, L2, L3)}
% \label{tab:factor-loadings}
% \begin{adjustbox}{max width=\linewidth}
% \begin{tabular}{lccc}
% \toprule
% \textbf{Variable} & \textbf{L1} & \textbf{L2} & \textbf{L3} \\
% \midrule
% SDS (social)             & 1.000 & 0.000 & 0.000 \\
% BPI on average            & -0.010 & 1.000& 0.000 \\
% Treatment satisfaction & 0.138 & -0.090 & 1.000\\
% Family social          & 0.138 & -0.040 & -0.466 \\
% Pain management (number of drugs)      & -0.164 & 0.395 & -0.446 \\
% Confidence in abstinence  & 0.007 & 0.173 & -0.610 \\
% SOWS                    & 0.047 & 0.209 & -0.996 \\
% SF12 (Physical Score)               & -0.050 & -0.559 & 0.071 \\
% SF12 (Mental Score)               & -0.188 & 0.002 & 1.232 \\
% SDS (work)               & 0.626 & 0.038 & -0.159 \\
% BPI at its best               & 0.090 & 0.766 & 0.120 \\
% K6                      & 0.212 & 0.051 & -1.252 \\
% BDI                     & -0.362 & 0.063 & -2.611 \\
% SDS (family)             & 0.739 & 0.037 & -0.324 \\
% BPI at its worst              & -0.048 & 0.891 & -0.060 \\
% \bottomrule
% \end{tabular}
% \end{adjustbox}
% \end{table}

\begin{figure}[tb]
    \centering
    \includegraphics[width=\linewidth]{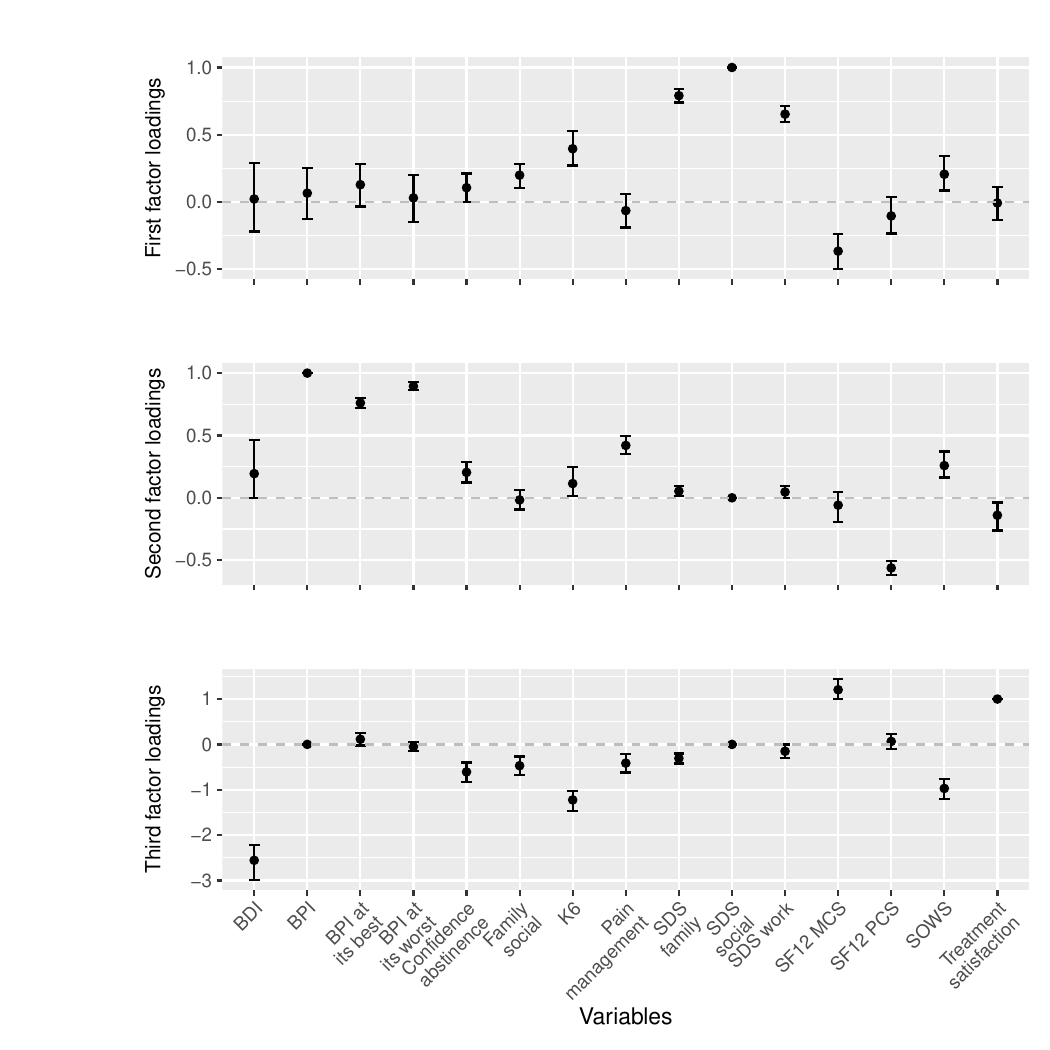}
    \caption{Estimated factor loading matrix $\mb{B}$ using BDCFM with 4 subgroups and 3 factors on the OUD recovery data. Posterior means are indicated with dots and 95\% credible intervals are indicated with vertical lines.}
    \label{fig:roadB}
\end{figure}
% functional impairment, physical, mental?
%We first look at the estimated factor loadings matrix to identify the three main dimensions of recovery by observing the associations between variables and factors. 
Figure \ref{fig:roadB} shows the posterior means and $95\%$ credible intervals of the factor loading matrix $\mb{B}$. As can be seen from the figure, the 95\% credible intervals of most factor loadings are very narrow, indicating high precision in their estimation. 
Based on the estimated loadings, we see that SDS family, SDS social, and SDS work are strongly related to the first factor. BPI on average, BPI at its best, BPI at its worst, SF12 physical score, and number of drugs used for pain management are strongly related to the second factor. Finally, BDI, K6, SF12 mental score, and confidence in abstinence strongly relate to the third factor.

\begin{table}[ht]
\centering
\caption{Summary of variables used in BDCFM with 4 subgroups and 3 factors on the OUD
recovery data separated by subgroup.  All data were standardized prior to summarization, and means and standard deviations are reported.}
\label{tab:cluster-wisemean}
\begin{adjustbox}{max width=\linewidth}
\begin{tabular}{lcccc}
%\toprule
\hline
 & \textbf{Subgroup 1} & \textbf{Subgroup 2} & \textbf{Subgroup 3} & \textbf{Subgroup 4} \\
\hline
\textbf{n} & 253 & 299 & 135 & 321 \\
\hline
% % \textbf{Dimension 1: Pain and Physical Health} & & & & \\
% BPI & -1.05 (0.00) & 0.34 (0.82) & 0.58 (0.98) & -0.02 (0.82) \\
% BPI at its worst  & -1.08 (0.36) & 0.40 (0.82) & 0.50 (0.88) & 0.07 (0.88) \\
% SF12 (Physical Score)  & 0.66 (0.59) & -0.14 (0.94) & -0.42 (1.11) & -0.03 (0.88) \\
% Pain management (number of drugs)  & -0.59 (0.53) & 0.15 (0.99) & 0.33 (1.14) & 0.08 (0.93) \\
% % \midrule
% % \textbf{Dimension 2: Psychological Distress and Mental Health} & & & & \\
% SF12 (Mental Score)  & 0.58 (0.60) & 0.47 (0.71) & -0.99 (0.86) & 0.00 (0.82) \\
% BDI  & -0.65 (0.18) & -0.37 (0.46) & 1.10 (1.15) & -0.27 (0.44) \\
% K6  & -0.68 (0.50) & -0.46 (0.63) & 1.08 (0.90) & -0.01 (0.63) \\
% Family social  & -0.24 (0.40) & -0.16 (0.85) & 0.41 (1.43) & -0.08 (0.71) \\
% Treatment satisfaction  & 0.23 (0.89) & 0.22 (0.83) & -0.40 (1.15) & -0.07 (0.94) \\
% Confidence in abstinence  & -0.50 (0.75) & -0.08 (0.92) & 0.50 (1.06) & 0.05 (0.96) \\
% % \midrule
% % \textbf{Dimension 3: Symptom Persistence} & & & & \\
% SOWS  & -0.47 (0.39) & -0.31 (0.60) & 0.77 (1.36) & -0.06 (0.62) \\
% BPI at its best  & -0.78 (0.16) & 0.18 (0.96) & 0.52 (1.13) & -0.03 (0.83) \\
% % \midrule
% % \textbf{Dimension 4: Social and Functional Impairment} & & & & \\
% SDS (social)  & -0.69 (0.00) & -0.69 (0.00) & 1.28 (0.90) & 0.07 (0.34) \\
% SDS (work)  & -0.54 (0.07) & -0.50 (0.27) & 0.99 (1.28) & 0.03 (0.64) \\
% SDS (family)  & -0.69 (0.15) & -0.61 (0.36) & 1.19 (0.94) & 0.10 (0.53) \\
          
SDS (social) &  -0.69 (0.00)&  -0.69 (0.00)&   0.37 (0.70) &  1.03 (1.00) \\
BPI &  -1.05 (0.00) &  0.33 (0.84) &  0.21 (0.99) &  0.43 (0.95)\\     
Treatment satisfaction (mean (SD))  &        0.24 (0.89) &  0.21 (0.83)&  -0.18 (0.97)&  -0.31 (1.14)\\  
Family social &   -0.24 (0.40) & -0.18 (0.77) & -0.08 (0.77) &  0.39 (1.41)\\    
Pain management (number of drugs) &   -0.60 (0.52) &  0.12 (0.98)&   0.03 (0.84)&   0.34 (1.15)\\    
Confidence in abstinence  &  -0.50 (0.75) & -0.10 (0.91)&  -0.02 (0.92) &  0.50 (1.07)\\ 
SOWS  &    -0.47 (0.38) & -0.33 (0.58) & -0.05 (0.67) &  0.70 (1.33)  \\   
SF12 (Physical Score) &    0.65 (0.60)&  -0.11 (0.95)&  -0.16 (0.95) & -0.34 (1.08) \\   
SF12 (Mental Score) & 0.59 (0.60) &  0.50 (0.69) &  0.10 (0.83) & -0.97 (0.82) \\   
SDS (work)   &   -0.54 (0.07)&  -0.50 (0.27)&   0.23 (0.85) &  0.80 (1.28) \\    
BPI at its best &  -0.78 (0.16) &  0.17 (0.97)&   0.23 (1.07) &  0.36 (1.06) \\   
K6  & -0.69 (0.49)&  -0.49 (0.60) &  0.01 (0.74) &  1.00 (0.89)  \\  
BDI  &  -0.66 (0.11)&  -0.42 (0.37)&  -0.50 (0.21) &  1.13 (1.05) \\  
SDS (family) &   -0.69 (0.15) & -0.62 (0.35)&   0.30 (0.79) &  1.00 (0.99) \\    
BPI at its worst &   -1.07 (0.36) &  0.38 (0.84) &  0.16 (0.94) &  0.42 (0.89) \\    \hline  
Time 1    &        62      &      79     &       27     &       84         \\   
Time 2     &       63      &      67    &        37      &      85         \\   
Time 3   &         63        &    77      &      32      &      80        \\    
Time 4      &      65       &     76    &        39     &       72     \\
\hline
\end{tabular}
\end{adjustbox}
\end{table}

\begin{figure}[tb]
    \centering
    \includegraphics[width=\linewidth]{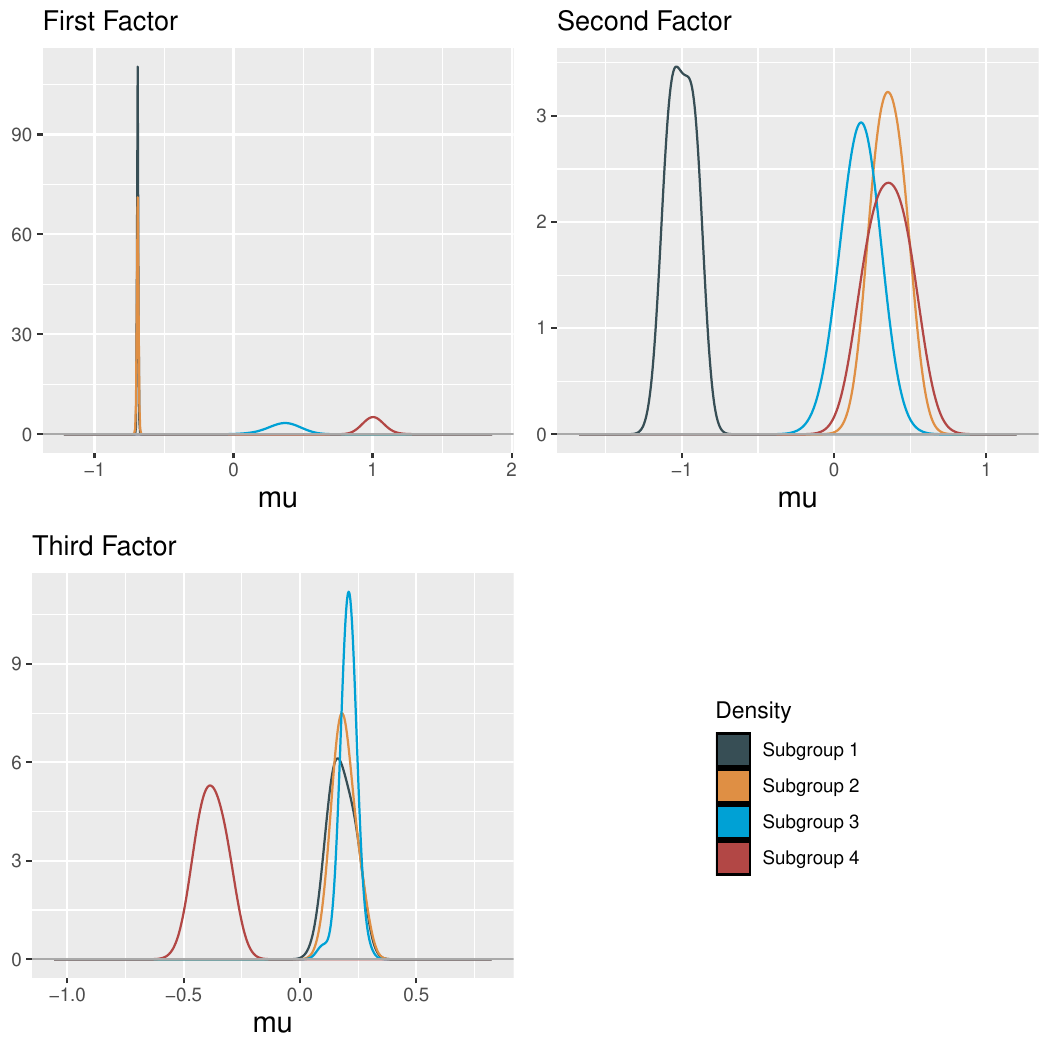}
    \caption{Estimated posterior density of $\mb{\mu}$ using BDCFM with 4 subgroups and 3 factors on the OUD recovery data.  Each panel represents the estimated posterior means of each cluster for one  factor. }
    \label{fig:roadmu}
\end{figure}

  Figure \ref{fig:roadmu} shows the posterior densities of the subgroup means $\BM{\mu}_g$ for $g=1,\dots, 4$. The first factor differentiates subgroups 1 and 2 from subgroups 3 and 4. The second factor differentiates subgroup 1 from all other subgroups. The third factor differentiates subgroup 4 from all other subgroups. The characteristic differences between subgroups are also replicated in the original variable space (Table \ref{tab:cluster-wisemean}).
 
%The first factor corresponds to variables related to functional impairment, with smaller values corresponding to less impairment. For example, for the first factor, the density for subgroups 1 and 2 is centered around -0.7, which means the subjects in these subgroups tend to have lower functional impairment. The second factor covers pain and physical health-related variables, with lower values corresponding to lower pain and higher physical health. For the second factor, the posterior densities of the mean for subgroups 2, 3, and 4 are wider and more to the right than that for subgroup 1. Therefore, subjects in subgroup 1 seem to be doing better in terms of pain and physical health. The third factor primarily covers psychological and mental health-related variables, with lower values corresponding to lower psychological and mental health. For the third factor, the density is wider, peaking around -0.35 for subgroup 4. For subgroups 1, 2, and 3, the mean of the third factor has a considerable overlap with no significant differences among these three subgroups. 

% \begin{figure}[H]
%     \centering
%     \includegraphics[width=\linewidth]{proj1/recover-data/Q_heatmap_g4l3s252times4.pdf}
%     \caption{RECOVER data -- heatmap of the transition probabilities for the 4-cluster-3-factor BDCFM.}
%     \label{fig:roadtransprob}
% \end{figure}

\begin{figure}[tb]
    \centering
    % First Subfigure
    \subfloat[]{
        \includegraphics[width=0.45\linewidth]{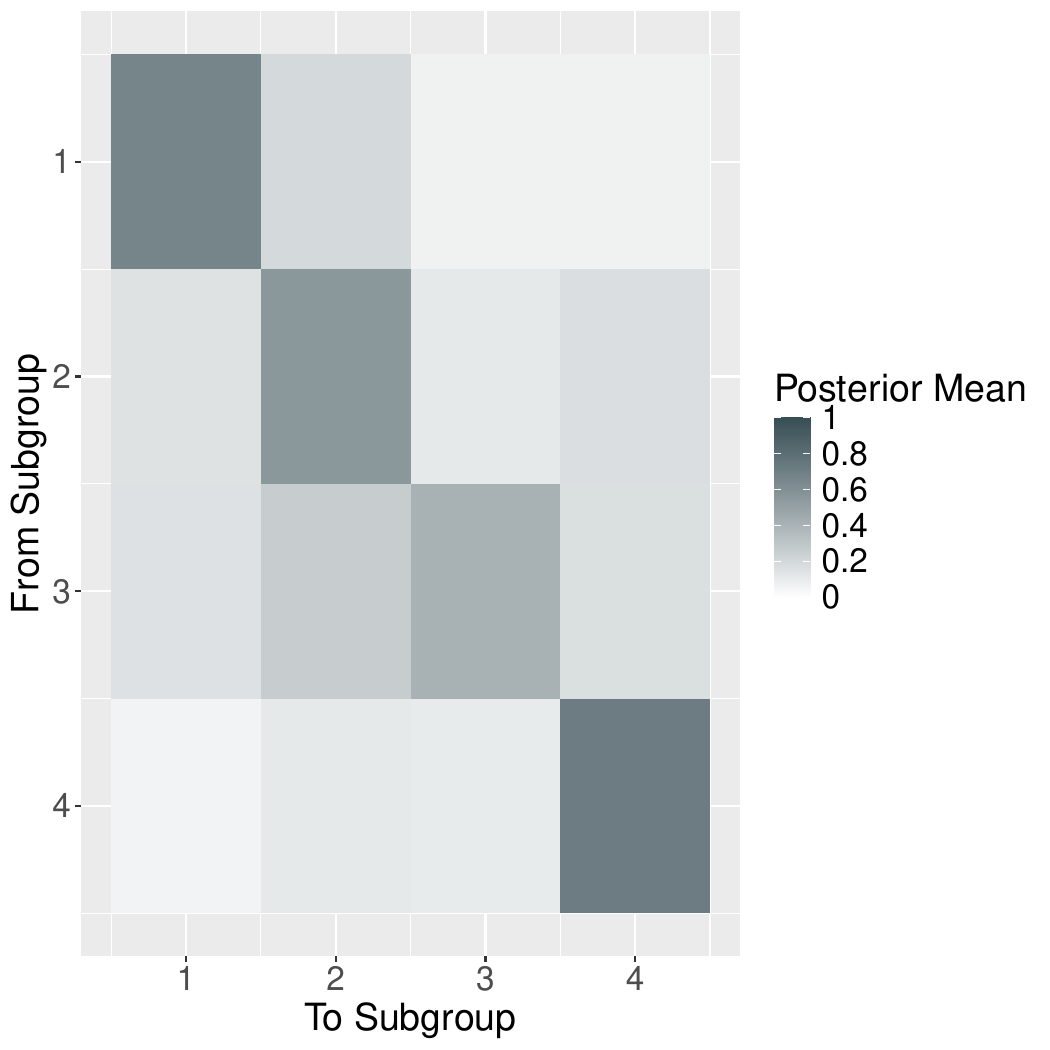} % Replace with your image
        \label{fig:figure3}
    }
    \hfill
    % Second Subfigure
    \subfloat[]{
        \includegraphics[width=0.45\linewidth]{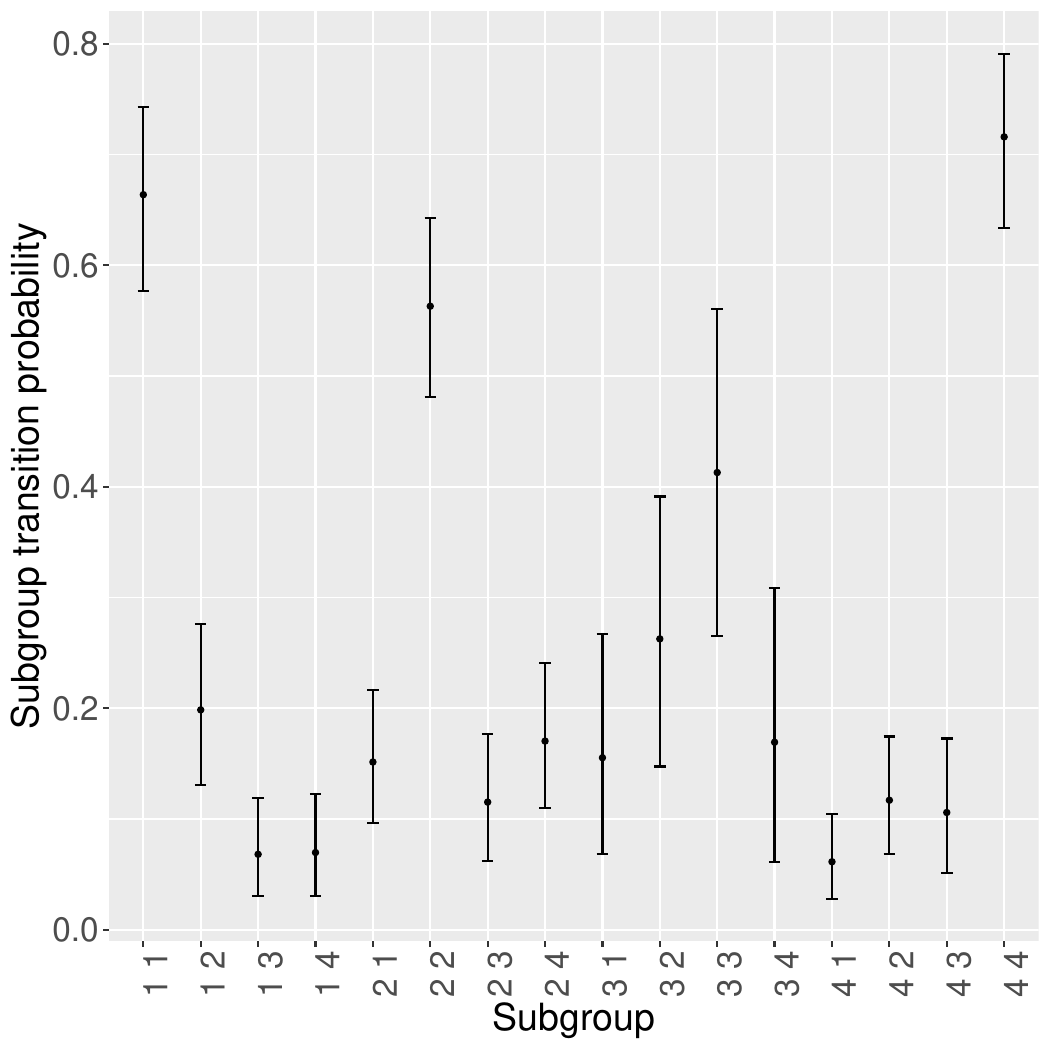} % Replace with your image
        \label{fig:figure4}
    }
    \caption{Estimated transition probability matrix $\mb{Q}$ using BDCFM with 4 subgroups and 3 factors on the OUD recovery data. (a) Heatmap of the posterior mean of transition probabilities. (b) Posterior mean and 95\% credible interval of subgroup transition probabilities. On the x-axis, the two digit notation represents the subgroup transitions. Specifically, the first digit is the subgroup  the subject is transitioning from, and the second  digit is the subgroup  the subject is transitioning to.}
    \label{fig:roadtransprob}
\end{figure}

The estimated initial probabilities of subgroup assignment for the subjects are 0.24, 0.32, 0.10, and 0.34 for subgroups 1 through 4, respectively. 
Moreover, BDCFM estimates the  subgroup transition probabilities $q_{jg}$ for $j=1,\dots, 4$ and $g=1,\dots, 4$ among the four subgroups (Figure \ref{fig:figure3}), and provides the $95\%$ credible interval for each of these transition probabilities (Figure \ref{fig:figure4}). Note that
the estimated probabilities of staying in the same group are 0.66, 0.56, 0.41, and 0.72 for subgroups 1, 2, 3, and 4, respectively.

\begin{figure}[tb]
    \centering
    \includegraphics[width=\linewidth]{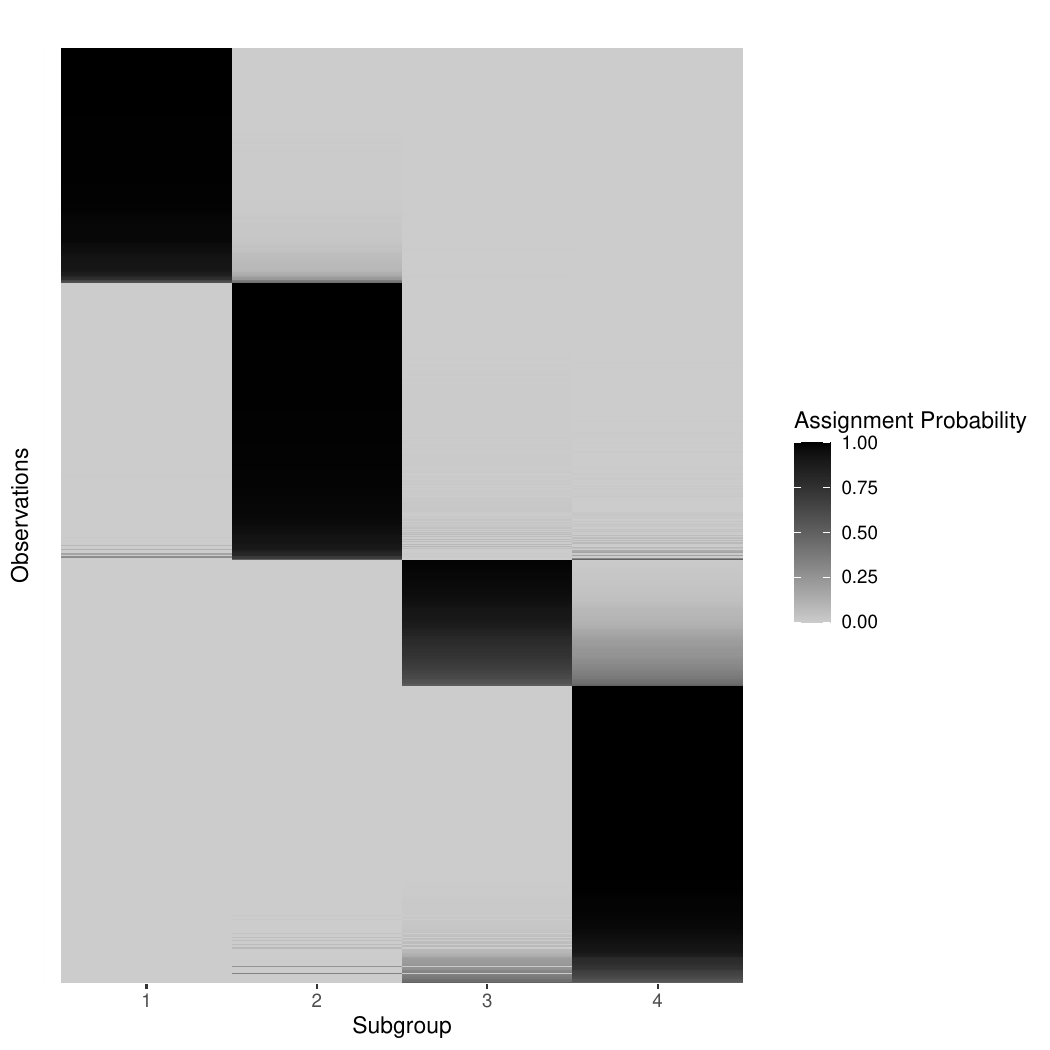}
    \caption{Subgroup assignment probabilities using BDCFM with 4 subgroups and 3 factors on the OUD recovery data. The heatmap includes subgroup assignment probabilities for $S=252$ subjects over $T=4$ time points. Subjects are ordered based on the subgroup with highest posterior probability. The shaded value represent the subgroup assignment probability.}
    \label{fig:roadZ}
\end{figure}

\begin{figure}[tb]
    \centering
    \includegraphics[width=\linewidth]{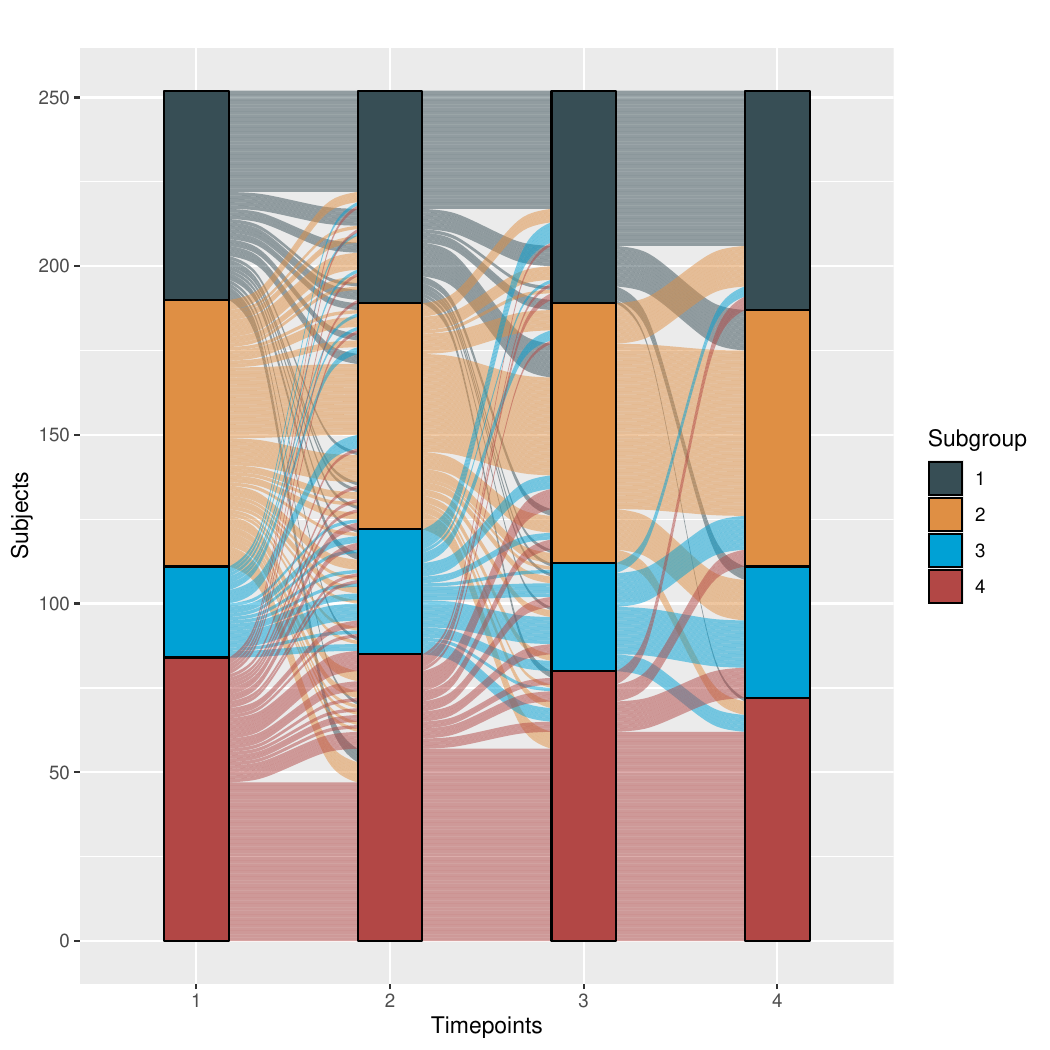}
    \caption{Posterior modes of $Z_{it}$ using BDCFM with 4 subgroups and 3 factors
on the OUD recovery data. The alluvial plot indicates the subgroup assignment at each time point and the subject-level transitions among subgroups over time. }
    \label{fig:roadCA}
\end{figure}

In addition to estimating the factor loading matrix, cluster means, and transition probabilities, BDCFM also estimates cluster assignments for each subject at each time point with uncertainty quantification (Figure \ref{fig:roadZ}). When considering cluster assignments over time (Figure \ref{fig:roadCA}), we observe the dynamic nature of OUD recovery.

% provide specific estimates.

% \begin{figure}[H]
%     \centering
%     \includegraphics[width=0.7\linewidth]{proj1/recover-data/sigma2_CI_g4l3s252times4.pdf}
%     \caption{RECOVER data -- uniqueness for the 4-subgroup-3-factor using BDCFM: posterior mean (black dot) and 95\% credible interval (vertical line).}
%     \label{fig:roadsigma}
    
% \end{figure}

% Figure \ref{fig:roadsigma} shows the posterior mean (black dot) and 95\% credible interval (vertical line) of the uniquenesses. Uniqueness represents the proportion of variance in a variable that is not explained by the common factors. The posterior means of uniquenesses for BPI, BDI, and SDS social are all very small, almost close to 0, and the credible intervals are so narrow that they are hidden under the posterior mean black dot. However, the uniqueness of variables about treatment satisfaction, family social relationship, confidence in abstinence, and number of drugs used for pain management are fairly high. This might be because a large part of its variance is specific to itself and not shared with the other variables. It is less influenced by the latent factors. 

\section{Discussion}
\label{sec:discussion}
 Here we have introduced our novel BDCFM framework that
 combines HMMs and factor models to capture the temporal dynamics in multivariate longitudinal data. BDCFM provides cluster assignments for each subject at each time point, while permitting subjects to transition among clusters over time.
  %A BDCFM analysis of multivariate longitudinal data finds the latent clusters and factors. 
  %To ameliorate the issue of label switching, we use an empirical Bayes approach based on a preliminary factor analysis combined with k-means clustering. 
  We have developed a Gibbs sampler to explore the posterior distribution of the parameters, which allows our BDCFM framework to provide both estimates of the unknown quantities and uncertainty quantification.

%properties of the method 
We evaluated the performance of our BDCFM framework through a  simulated dataset. 
BDCFM accurately estimated the parameters. 
In addition, 
 BDCFM properly quantified the uncertainty and was well calibrated. Specifically, the 95\% credible intervals contained the true values of the parameters more than 95\% of the time.
Lastly, BDCFM correctly assigned subjects to clusters at each time point with a very low misclassification rate. 

We have demonstrated the utility of our BDCFM framework with an analysis of a dataset on OUD recovery. This dataset had $S=252$ subjects across $T=4$ time points, and the BDCFM framework completed this analysis in 6.2 minutes. Lastly, we have created an R package that will be available on CRAN upon publication.

 %This, in turn, may help recovery professionals design more customized treatment plans for each of the subgroups. 
 
 % Individuals in the high-functioning subgroup have the best recovery profile and mental health and, thus, require less psychological support. However, individuals in the low-functioning subgroup show they not only struggle with the 3 dimensions of the recovery process we identified but also have dissatisfaction with their treatment and have serious conflict in their family/social relationships. This identification can now help the relevant professional to have a different approach to treating individuals in this subgroup. However, we need to consider that this data only included individuals who completed all five time points during the 24 months of the RECOVER study.
 
 There are several promising directions for future research. One such direction is the  development of  methods for the choice of the number of factors and the number of clusters. Another possible future direction  is the inclusion of regressors in the BDCFM framework. This may be used to estimate the impact of regressors such as changes in individual-level psychosocial factors on the probability of transition among clusters.  
 % With this new addition, we can observe if the perceived treatment effectiveness will increase the probability of transitions from a low-functioning to a high-functioning subgroup. 

\bibliographystyle{natbib}
\bibliography{proj1}

\begin{thebibliography}{31}
\providecommand{\natexlab}[1]{#1}
\providecommand{\url}[1]{\texttt{#1}}
\expandafter\ifx\csname urlstyle\endcsname\relax
  \providecommand{\doi}[1]{doi: #1}\else
  \providecommand{\doi}{doi: \begingroup \urlstyle{rm}\Url}\fi

\bibitem[Craft et~al.(2023)Craft, Shin, Tegge, Keith, Athamneh, Stein,
  Ferreira, Chilcoat, Le~Moigne, DeVeaugh-Geiss, and Bickel]{craft2023longterm}
William~H. Craft, Hwasoo Shin, Allison~N. Tegge, Diana~R. Keith, Liqa~N.
  Athamneh, Jeffrey~S. Stein, Marco A.~R. Ferreira, Howard~D. Chilcoat, Anne
  Le~Moigne, Angela DeVeaugh-Geiss, and Warren~K. Bickel.
\newblock Long-term recovery from opioid use disorder: recovery subgroups,
  transition states and their association with substance use, treatment and
  quality of life.
\newblock \emph{Addiction}, 118:\penalty0 890--900, 2023.
\newblock \doi{10.1111/adb.13291}.

\bibitem[Yeung and Ruzzo(2001)]{yeung2001principal}
Ka~Yee Yeung and Walter~L. Ruzzo.
\newblock Principal component analysis for clustering gene expression data.
\newblock \emph{Bioinformatics}, 17\penalty0 (9):\penalty0 763--774, 2001.

\bibitem[Hastie et~al.(2009)Hastie, Tibshirani, and
  Friedman]{hastie2009elements}
Trevor Hastie, Robert Tibshirani, and Jerome Friedman.
\newblock \emph{The elements of statistical learning: data mining, inference,
  and prediction}.
\newblock Springer, second edition, 2009.

\bibitem[Lopes and West(2004)]{LopesWest2004}
H.~Lopes and M.~West.
\newblock Bayesian model assessment in factor analysis.
\newblock \emph{Statistica Sinica}, 14:\penalty0 41--67, 2004.

\bibitem[Shin and Ferreira(2023)]{shin2023dynamic}
Hwasoo Shin and Marco~A.R. Ferreira.
\newblock Dynamic {ICAR} spatiotemporal {F}actor {M}odels.
\newblock \emph{Spatial Statistics}, 56:\penalty0 100763, 2023.
\newblock \doi{10.1016/j.spasta.2023.100763}.
\newblock URL \url{https://doi.org/10.1016/j.spasta.2023.100763}.

\bibitem[{Fr\"uhwirth-Schnatter}(2006)]{fruhwirth:2006}
S.~{Fr\"uhwirth-Schnatter}.
\newblock \emph{Finite Mixture and {M}arkov Switching Models}.
\newblock Springer-Verlag, 2006.

\bibitem[Fruhwirth-Schnatter et~al.(2019)Fruhwirth-Schnatter, Celeux, and
  Robert]{fruhwirth2019handbook}
Sylvia Fruhwirth-Schnatter, Gilles Celeux, and Christian~P Robert.
\newblock \emph{Handbook of Mixture Analysis}.
\newblock CRC press, 2019.

\bibitem[Shin et~al.(2025)Shin, Ferreira, and
  Tegge]{shin2025bayesianclusteringfactormodels}
Hwasoo Shin, Marco A.~R. Ferreira, and Allison~N. Tegge.
\newblock Bayesian clustering factor models, 2025.
\newblock URL \url{https://arxiv.org/abs/2505.05280}.

\bibitem[Gelfand and Smith(1990{\natexlab{a}})]{gelfand:smith:1990}
Alan~E Gelfand and Adrian F~M Smith.
\newblock Sampling-based approaches to calculating marginal densities.
\newblock \emph{Journal of the American Statistical Association}, 85\penalty0
  (410):\penalty0 398--409, 1990{\natexlab{a}}.

\bibitem[Robert and Casella(2005)]{robe:case:2005}
C.~P. Robert and G.~Casella.
\newblock \emph{Monte {C}arlo Statistical Methods}.
\newblock Springer-Verlag, New York, 2nd edition, 2005.

\bibitem[Gamerman and Lopes(2006)]{game:lope:2006}
Dani Gamerman and Hedibert~F. Lopes.
\newblock \emph{Markov Chain {M}onte {C}arlo: Stochastic Simulation for
  {B}ayesian Inference}.
\newblock Chapman and Hall/CRC, Boca Raton, FL, 2nd edition, 2006.

\bibitem[Kass and Wasserman(1995)]{kass:1995}
Robert~E Kass and Larry Wasserman.
\newblock A reference {B}ayesian test for nested hypotheses and its
  relationship to the {S}chwarz criterion.
\newblock \emph{Journal of the American Statistical Association}, 90\penalty0
  (431):\penalty0 928--934, 1995.

\bibitem[Steele and Raftery(2010)]{steele2010performance}
Russell~J Steele and Adrian~E Raftery.
\newblock Performance of {B}ayesian model selection criteria for {G}aussian
  mixture models.
\newblock In M.~Chen, Peter Muller, Dongchu Sun, Keying Ye, and Dipak Dey,
  editors, \emph{Frontiers of Statistical Decision Making and Bayesian
  Analysis}, pages 113--130. Springer, 2010.

\bibitem[Fokou{\'e} and Titterington(2003)]{fokoue2003mixtures}
Ernest Fokou{\'e} and DM~Titterington.
\newblock Mixtures of factor analysers. bayesian estimation and inference by
  stochastic simulation.
\newblock \emph{Machine Learning}, 50:\penalty0 73--94, 2003.

\bibitem[Papastamoulis(2018)]{papastamoulis2018overfitting}
Panagiotis Papastamoulis.
\newblock Overfitting {B}ayesian mixtures of factor analyzers with an unknown
  number of components.
\newblock \emph{Computational Statistics \& Data Analysis}, 124:\penalty0
  220--234, 2018.

\bibitem[Chandra et~al.(2023)Chandra, Canale, and Dunson]{chandra2023escaping}
Noirrit~Kiran Chandra, Antonio Canale, and David~B Dunson.
\newblock Escaping the curse of dimensionality in {B}ayesian model-based
  clustering.
\newblock \emph{Journal of Machine Learning Research}, 24\penalty0
  (144):\penalty0 1--42, 2023.

\bibitem[Ghilotti et~al.(2025)Ghilotti, Beraha, and
  Guglielmi]{ghilotti2025bayesian}
Lorenzo Ghilotti, Mario Beraha, and Alessandra Guglielmi.
\newblock Bayesian clustering of high-dimensional data via latent repulsive
  mixtures.
\newblock \emph{Biometrika}, 112\penalty0 (2):\penalty0 asae059, 2025.

\bibitem[Geweke and Zhou(1996)]{geweke:zhou:1996}
John Geweke and Guofu Zhou.
\newblock Measuring the price of the arbitrage pricing theory.
\newblock \emph{The Review of Financial Studies}, 9\penalty0 (2):\penalty0
  557--587, 1996.
\newblock ISSN 08939454, 14657368.

\bibitem[Aguilar and West(1999)]{aguilar1999bayesian}
Omar Aguilar and Mike West.
\newblock Bayesian dynamic factor models and portfolio allocation.
\newblock \emph{Journal of Business \& Economic Statistics}, 18\penalty0
  (3):\penalty0 338--357, 1999.
\newblock \doi{10.1080/07350015.2000.10524868}.

\bibitem[Prado et~al.(2021)Prado, Ferreira, and West]{PradoFerreiraWest2021}
R.~Prado, M.~A.~R. Ferreira, and M.~West.
\newblock \emph{Time Series: Modeling, Computation, and Inference. {S}econd
  Edition.}
\newblock Boca Raton: CRC Press, 2021.

\bibitem[Gelfand and Smith(1990{\natexlab{b}})]{gelfand1990sampling}
Alan~E. Gelfand and Adrian F.~M. Smith.
\newblock Sampling-based approaches to calculating marginal densities.
\newblock \emph{Journal of the American Statistical Association}, 85\penalty0
  (410):\penalty0 398--409, 1990{\natexlab{b}}.
\newblock \doi{10.2307/2289776}.

\bibitem[Sheehan et~al.(1998)Sheehan, Lecrubier, Sheehan, Amorim, Janavs,
  Weiller, Hergueta, Baker, and Dunbar]{sheehan1998mini}
David~V. Sheehan, Yves Lecrubier, Kathleen~Harnett Sheehan, Pedro Amorim,
  Judith Janavs, Edouard Weiller, T.~Hergueta, R.~Baker, and Graeme~C. Dunbar.
\newblock The mini-international neuropsychiatric interview (mini): The
  development and validation of a structured diagnostic psychiatric interview
  for {DSM-IV} and {ICD-10}.
\newblock \emph{Journal of Clinical Psychiatry}, 59\penalty0 (Suppl
  20):\penalty0 22--33, 1998.

\bibitem[Kessler et~al.(2003)Kessler, Barker, Colpe, Epstein, Gfroerer, Hiripi,
  Howes, Normand, Manderscheid, Walters, and Zaslavsky]{kessler2003screening}
Ronald~C. Kessler, Patricia~R. Barker, Lisa~J. Colpe, Joan~F. Epstein,
  Joseph~C. Gfroerer, Eva Hiripi, Mary~J. Howes, Sharon-Lise~T. Normand,
  Ronald~W. Manderscheid, Ellen~E. Walters, and Alan~M. Zaslavsky.
\newblock Screening for serious mental illness in the general population.
\newblock \emph{Archives of General Psychiatry}, 60\penalty0 (2):\penalty0
  184--189, 2003.
\newblock \doi{10.1001/archpsyc.60.2.184}.

\bibitem[Handelsman et~al.(1987)Handelsman, Cochrane, Aronson, Ness,
  Rubinstein, and Kanof]{handelsman1987two}
Lester Handelsman, Kenneth~J. Cochrane, Marvin~J. Aronson, Ronald Ness,
  Kenneth~J. Rubinstein, and Philip~D. Kanof.
\newblock Two new rating scales for opiate withdrawal.
\newblock \emph{American Journal of Drug and Alcohol Abuse}, 13:\penalty0
  293--308, 1987.
\newblock \doi{10.3109/00952998709001515}.

\bibitem[Mendoza et~al.(2006)Mendoza, Mayne, Rublee, and
  Cleeland]{mendoza2006reliability}
Tito Mendoza, Tracy Mayne, Diane Rublee, and Charles Cleeland.
\newblock Reliability and validity of a modified brief pain inventory short
  form in patients with osteoarthritis.
\newblock \emph{European Journal of Pain}, 10:\penalty0 353--356, 2006.
\newblock \doi{10.1016/j.ejpain.2005.06.002}.

\bibitem[Beck et~al.(1996)Beck, Steer, and Brown]{beck1996bdi}
Aaron~T. Beck, Robert~A. Steer, and Gregory~K. Brown.
\newblock \emph{Manual for the Beck Depression Inventory-II}.
\newblock Psychological Corporation, San Antonio, TX, 1996.

\bibitem[Ware et~al.(1996)Ware, Kosinski, and Keller]{ware1996sf12}
John~Jr Ware, Mark Kosinski, and Susan~D. Keller.
\newblock A 12-item short-form health survey: construction of scales and
  preliminary tests of reliability and validity.
\newblock \emph{Medical Care}, 34\penalty0 (3):\penalty0 220--233, 1996.
\newblock \doi{10.1097/00005650-199603000-00003}.

\bibitem[Revelle(2024)]{Rpackagefa}
W.~Revelle.
\newblock \emph{psych: Procedures for Psychological, Psychometric, and
  Personality Research}.
\newblock Northwestern University, Evanston, Illinois, 2024.
\newblock URL \url{https://CRAN.R-project.org/package=psych}.
\newblock R package version 2.4.3.

\bibitem[Thomson(1951)]{thomson1951}
G.~H. Thomson.
\newblock \emph{The Factorial Analysis of Human Ability. {F}ifth {E}dition.}
\newblock University of London Press, 1951.

\bibitem[Tibshirani et~al.(2001)Tibshirani, Walther, and
  Hastie]{tibshirani2001estimating}
Robert Tibshirani, Guenther Walther, and Trevor Hastie.
\newblock Estimating the number of clusters in a data set via the gap
  statistic.
\newblock \emph{Journal of the Royal Statistical Society: Series B (Statistical
  Methodology)}, 63\penalty0 (2):\penalty0 411--423, 2001.
\newblock \doi{10.1111/1467-9868.00293}.

\bibitem[Trefethen and Bau~III(1997)]{trefethen1997numerical}
Lloyd~N. Trefethen and David Bau~III.
\newblock \emph{Numerical Linear Algebra}.
\newblock Society for Industrial and Applied Mathematics (SIAM), Philadelphia,
  PA, 1997.
\newblock ISBN 978-0-898713-61-9.

\end{thebibliography}

\newpage

\section{Appendix}

\subsection{Empirical Bayes specification of priors for cluster parameters}\label{sec:specificationofpriors}

BDCFM inherits some computational challenges from hidden Markov models. 
Specifically, if the prior distribution on the cluster parameters is the same for each cluster, then the posterior distribution has several modes corresponding to different label orders of the clusters. 
In that case, MCMC implementations may suffer from label switching, where two or more clusters may switch labels. 
To prevent label switching, similarly to \cite{steele2010performance}, we implement a prior specification for the parameters of each cluster based on empirical Bayes. 

The empirical Bayes procedure combines factor analysis with $k$-means clustering. 
First, we stack all transposed observed vectors $\mb{y}_{it}'$ by row into a matrix $\mb{Y}$ with $n=S\times T$ rows and $R$ columns. 
After that, we obtain an estimate $\Tilde{\BM{B}}$ of the factor loadings matrix using the function \verb|fa| from the \verb|R| package \verb|psych| \citep{Rpackagefa}. 
While using \verb|fa|, we set the method argument to "minimum residual" and the rotate argument to "none". 
After that, we compute a matrix $\BM{M}$ that transforms $\Tilde{\BM{B}}$ into a matrix $\widehat{\BM{B}} = \Tilde{\BM{B}}\BM{M}$ that satisfies the hierarchical structural constraint. 
From the output of the function \verb|fa|, we also obtain a preliminary estimate $\widehat{\BM{V}} = diag(\widehat{\sigma}^2_1,\dots,\widehat{\sigma}^2_R)$, where $\widehat{\sigma}^2_r$ is the estimated uniqueness that is equal to one minus the estimated communality of the $r$-th variable. 

To compute preliminary estimates of the vectors of common factors, our empirical Bayes approach uses the weighted least squares estimate \citep{thomson1951}
% include text from previous version back.  
$
\widehat{\BM{X}} = (\widehat{\BM{B}}'\widehat{\BM{V}}^{-1}\widehat{\BM{B}})^{-1}\widehat{\BM{B}}'\widehat{\BM{V}}^{-1}\mb{Y},
$
where each row of $\widehat{\BM{X}}$ corresponds to an estimated vector of common factors for a given subject at a specific time point. 
Note that the computation of $\widehat{\BM{X}}$ does not perform any shrinkage towards a common mean vector, which would be undesirable in the case when the vectors of common factors come from a mixture of Gaussians as assumed by BDCFM. 
We then cluster the estimated vectors of common factors with $k$-means clustering \citep{tibshirani2001estimating}. 
After that, we order the clusters in decreasing order of cluster size. 
Let $\widetilde{\BM{\mu}}_g$ and $\widetilde{\BM{\Omega}}_g$, $g=1,\dots,G$, respectively be the sample mean vector and sample covariance matrix of the estimated vectors of common factors assigned to cluster $g$. 

To respect the constraint that the first cluster has a diagonal covariance matrix $\BM{\Omega}_1$, we compute the LDL decomposition \citep{trefethen1997numerical} of $\widetilde{\BM{\Omega}}_1$, that is, $\widetilde{\BM{\Omega}}_1 = \BM{L}_1 \BM{D}_1 \BM{L}_1'$, where $\BM{L}_1$ is a lower triangular matrix and $\BM{D}_1$ is a diagonal matrix. 
Then, we premultiply all preliminary estimates of vectors of common factors by $\BM{L}_1^{-1}$. 
As a result, the sample covariance matrix of the estimated vectors of common factors allocated to the first cluster becomes the diagonal matrix
$
    \BM{C}_1^{\mu} = \BM{L}_1^{-1}\widetilde{\BM{\Omega}}_1 \BM{L}_1^{-1'} = \BM{D}_1
$.
Denote by $s_{\omega_1}^2,\dots,s_{\omega_L}^2$ the diagonal elements of $\BM{C}_1^{\mu}$. 
Recall that the covariance matrix of the first cluster is $\BM{\Omega}_1=(\omega_1,\dots,\omega_L)$. 
Then, we assign for $\omega_l$ a prior $IG(n_{\omega_l}/2,n_{\omega_l}s_{\omega_l}^2/2)$, where $n_{\omega_l}/2 = 2$ implies that the mean of the $l$-th element $\omega_{l}$ is $s_{\omega_{l}}^2$. 

Further, the sample covariance matrix of the estimated vectors of common factors assigned to the $g$-th cluster becomes
$
    \BM{C}_g^{\mu} = \BM{L}_1^{-1} \widetilde{\BM{\Omega}}_g\BM{L}_1^{-1'}. 
$
Thus, we assume for the covariance matrix $\BM{\Omega}_g$ of the $g$-th cluster an inverse Wishart prior with parameters $\BM{S}_{\Omega_g} = \BM{C}_g^\mu$ and $n_{\Omega_g} = L + 2$, implying that the prior mean of $\BM{\Omega}_g$ is $\BM{S}_{\Omega_g}$. 
Finally, the mean vector $\BM{\mu}_g$ of the $g$-th cluster is assigned prior $N(\BM{m}_g^\mu, \BM{C}_g^\mu)$ where 
$
    \BM{m}_g^\mu = \BM{L}_1^{-1} \widetilde{\BM{\mu}}_g.
$

\subsection{Full conditional distributions}
\label{sec:fullconditionals.appendix}
\paragraph{Common factor $\BM{x}_{it}$}
If subject $i$ at time $t$ is assigned to cluster $g$ such that $Z_{it}=g$, then the full conditional density of $\BM{x}_{it}$ is
% \begin{equation*}
% \resizebox{.95\columnwidth}{!}{$
\begin{eqnarray*}
% \begin{autobreak}
p(\BM{x}_{it} \mid-) & \propto &p(\BM{x}_{it}) p(\BM{x}_{it} \mid \BM{\mu}_g, \BM{\Omega}_g, Z_{it}=g) \\
& \propto &\exp \left[-\frac{1}{2}\left\{\left(\BM{y}_{it}^{\prime}-\BM{x}_{it}^{\prime} \BM{B}^{\prime}) \BM{V}^{-1}(\BM{y}_{it}-\BM{B} \BM{x}_{it})+(\BM{x}_{it}^{\prime}-\BM{\mu}_g^{\prime}) \BM{\Omega}_g^{-1}(\BM{x}_{it}-\BM{\mu}_g\right)\right\}\right] \\
& \propto& \exp\left[-\frac{1}{2}\left(\BM{x}_{it}^{\prime} \BM{B}^{\prime} \BM{V}^{-1} \BM{B} \BM{x}_{it}-2 \BM{x}_{it}^{\prime} \BM{B}^{\prime} \BM{V}^{-1} \BM{y}_{it}+\BM{x}_{it}^{\prime} \BM{\Omega}_g^{-1} \BM{x}_{it}-2 \BM{x}_{it}^{\prime} \BM{\Omega}_g^{-1} \BM{\mu}_g\right)\right] \\
& \propto& \exp\left[-\frac{1}{2}\left\{\BM{x}_{it}^{\prime}
\left(\BM{B}^{\prime} \BM{V}^{-1} \BM{B}+\BM{\Omega}_g^{-1}\right) \BM{x}_{it}-2 \BM{x}_{it}^{\prime}\left(\BM{B}^{\prime} \BM{V}^{-1} \BM{y}_{it}+\BM{\Omega}_g^{-1} \BM{\mu}_g\right)\right\}\right].
% \end{autobreak}
\end{eqnarray*}
% $}
% \end{equation*}
Therefore, the full conditional distribution of $\BM{x}_{it}$ is $N((\BM{\Omega}_g^{-1}+\BM{B}^{\prime} \BM{V}^{-1} \BM{B})^{-1}(\BM{B}^{\prime} \BM{V}^{-1} \BM{y}_{it}+\BM{\Omega}_g^{-1} \BM{\mu}_g),(\BM{\Omega}_g^{-1}+\BM{B}^{\prime} \BM{V}^{-1} \BM{B})^{-1})$.

% $$
% N((B^{'}V^{-1}B+\Omega_j^{-1})^{-1}(B^{'}V^{-1}y_{it} + \Omega_j^{-1}\mu_j) ,(B^{'}V^{-1}B+\Omega_j^{-1})^{-1})
% $$

% Where $\Omega_j$ and $\mu_j$ are from the variance and mean parameters from the $j$-th cluster. $V$ is the variance structure from $\sigma^2$. 

\paragraph{Covariance matrix $\BM{\Omega}_g$}
%\subsubsection{Covariance matrix \texorpdfstring{$\BM{\Omega}_j$}{2}:}

Recall that the covariance matrix $\BM{\Omega}_1$ of the first cluster is diagonal, and  its $l$-th diagonal element, $l=1,...,L$, is $\omega_l$.
Then, the full conditional density of $\omega_l$ is
% \begin{equation*}
% \resizebox{.95\columnwidth}{!}{$
\begin{eqnarray*}
p(\omega_l|\cdot) & \propto& p(\omega_l) \prod_{t=1}^T  \prod_{i \in \mc{C}_{t1}} p(\BM{x}_{it}|\mu_1,\omega_l,Z_{it}=1)\\ \nonumber
& \propto& \omega_l ^{-\sum_{t=1}^Tn_{t1}/2-n_{\omega_l}/2-1} \exp \left[ -\frac{1}{2 \omega_l}\sum_{t=1}^T\sum_{i \in \mc{C}_{t1}} \left(x_{itl} - \mu_{1l}\right)^2 + n_{\omega_l} s_{\omega_l}^2\right], \nonumber
\end{eqnarray*}
% $}
% \end{equation*}
where $n_{t1}$ is the cardinality of cluster 1 at time $t$. 
Then, the full conditional distribution of $\omega_l$ is $IG\left[\left(n_{\omega_l}+\sum_{t=1}^Tn_{t1}\right)/2, \left\{\sum_{t=1}^T\sum_{i \in \mc{C}_{t1}}\left(x_{itl}-\mu_{1l}\right)^2+n_{\omega_l} s_{\omega_l} ^2\right\}/2\right] $. 

Now consider the covariance matrix of the $g$-th cluster $\BM{\Omega}_g$, $g>1$. 
Then, the full conditional density of $\BM{\Omega}_g$ is
\begin{eqnarray*}
% p(x_{it}|\mu_j\Omega_j,Z_i=j) & \propto (2\pi)^{-1/2} |\Omega_j|^{-1/2}exp(-\frac{1}{2}(x_{it}-\mu_j)^{T}\Omega_j^{-1}(x_{it}-\mu_j)^T)\\
% &= (2\pi)^{-1/2} |\Omega_j|^{-1/2}exp(-\frac{1}{2}tr(\Omega_j^{-1}(x_{it}-\mu_j)(x_{it}-\mu_j)^T)
% p(\Omega_j)= \frac{|\BM{S}_{\Omega_j}|^{n_{\Omega_j}/2}}{2^{n_{\Omega_j} F/2}\Gamma_F(n_{\Omega_j}/2)} |\Omega_j|^{-(n_{\Omega_j}+F+1)/2}exp(-\frac{1}{2}tr(\Omega_j^{-1}\BM{S}_{\Omega_j}))
    p(\BM{\Omega}_g|\cdot) & \propto & p(\BM{\Omega}_g) \prod_{t=1}^T\prod_{i \in \mc{C}_{tg}} p(\BM{x}_{it}|\BM{\mu}_g,\BM{\Omega}_g,Z_{it}=g)\\
    & \propto & |\BM{\Omega}_g|^{-(\sum_{t=1}^Tn_{tg}+n_{\Omega_g}+L+1)/2} \exp[-\frac{1}{2}\{tr(\BM{\Omega}_g^{-1} \sum_{t=1}^T\sum_{i \in \mc{C}_{tg}} (\BM{x}_{it} - \BM{\mu}_g)(\BM{x}_{it} - \BM{\mu}_g)')+tr(\BM{\Omega}_g^{-1}\BM{S}_{\Omega_g})\}]\\
    & \propto & |\BM{\Omega}_g|^{-(\sum_{t=1}^Tn_{tg}+n_{\Omega_g}+L+1)/2} \exp[-\frac{1}{2}\{tr(\BM{\Omega}_g^{-1} \{\sum_{t=1}^T\sum_{i \in \mc{C}_{tg}} (\BM{x}_{it} - \BM{\mu}_g)(\BM{x}_{it} - \BM{\mu}_g)'+\BM{S}_{\Omega_g}\})\}].
\end{eqnarray*}
Thus, the full conditional distribution of $\BM{\Omega}_g$ is inverse Wishart with shape parameter $\sum_{t=1}^Tn_{tg} + n_{\Omega_g}$ and scale parameter $\sum_{t=1}^T\sum_{i \in \mc{C}_{tg}}(\BM{x}_{it}-\BM{\mu}_g)(\BM{x}_{it}-\BM{\mu}_g)' + \BM{S}_{\Omega_g}$.

\paragraph{Cluster mean vector $\BM{\mu}_g$}
The full conditional density of $\BM{\mu}_g$ is
\begin{eqnarray*}
p(\BM{\mu}_g|\cdot) &\propto&  p(\BM{\mu}_g) \prod_{t=1}^T \prod_{i \in \mc{C}_{tg}} p(\BM{x}_{it}\mid \BM{\mu}_g, \BM{\Omega}_g, Z_{it} = g)\\
& \propto&  \exp [-\frac{1}{2} \sum_{t=1}^T\sum_{i \in \mc{C}_{tg}}(\BM{x}_{it}-\BM{\mu}_g)^{\prime} \BM{\Omega}_g^{-1}\BM{x}_{it}-\BM{\mu}_g)] \exp [-\frac{1}{2}(\BM{\mu}_g-\BM{m}_g^\mu)^{\prime} \BM{C}_g^{\mu-1}(\BM{\mu}_g-\BM{m}_g^\mu)] \\
& \propto&  \exp [-\frac{1}{2} \sum_{t=1}^T\sum_{i \in \mc{C}_{tg}}(\BM{\mu}_g^{\prime} \BM{\Omega}_g^{-1} \BM{\mu}_g-2 \BM{\mu}_g \BM{\Omega}_g^{-1} \BM{x}_{it})] \exp [-\frac{1}{2}(\BM{\mu}_g^{\prime} \BM{C}_g^{\mu-1} \BM{\mu}_g-2 \BM{\mu}_g \BM{C}_g^{\mu-1} \BM{m}_g^\mu)] \\
& \propto&  \exp [-\frac{1}{2} \BM{\mu}_g^{\prime}(\sum_{t=1}^Tn_{tg} \BM{\Omega}_g^{-1}+\BM{C}_g^{\mu-1}) \BM{\mu}_g-\BM{\mu}_g(\BM{\Omega}_g^{-1} \sum_{t=1}^T\sum_{i \in \mc{C}_{tg}} \BM{x}_{it}+\BM{C}_g^{\mu-1} \BM{m}_g^\mu)].
\end{eqnarray*}
% $}
% \end{equation*}

Therefore, the full conditional distribution of $\BM{\mu}_g$ is $N((\BM{C}_g^{\mu-1}+\sum_{t=1}^Tn_{tg} \BM{\Omega}_g^{-1})^{-1}(\BM{C}_g^{\mu-1} \BM{m}_g^\mu+ \BM{\Omega}_g^{-1} \sum_{t=1}^T \sum_{i \in \mc{C}_{tg}} \BM{x}_{it}),(\BM{C}_g^{\mu-1}+ \BM{\Omega}_g^{-1}\sum_{t=1}^Tn_{tg})^{-1})$.

\paragraph{Matrix of factor loadings $\mb{B}$}
Let $\BM{b}_{r.}$ be the $r$-th row of the factor loadings matrix $\BM{B}$. 
When $1<r<L$, the last $L-r+1$ elements of $\BM{b}_{r.}$ are fixed due to the hierarchical structural constraint. 
Specifically, the $r$-th element is 1, and the last $L-r$ elements are equal to 0. 
Let $\BM{b}_{r.}^*$ be a vector that contains the first $r-1$ elements of $\BM{b}_{r.}$. 
Recall that $\BM{\Upsilon}=diag(\tau_1,\ldots,\tau_L)$. 
Then, when $1<r<L$, the full conditional density of $\BM{b}_{r.}^*$ is
% \begin{equation*}
% \resizebox{.95\columnwidth}{!}{$
% \begin{eqnarray*}
% \begin{autobreak}
\begin{eqnarray*}
p(\BM{b}_{r.}^* \mid-)
& \propto &\exp \left\{-\frac{1}{2} \sum_{t=1}^T \sum_{i=1}^S(\BM{y}_{it}-\BM{B}\BM{x}_{it})^{\prime} \BM{V}^{-1}(\BM{y}_{it}-\BM{B} \BM{x}_{it})\right\} p(\BM{b}_{r.}^*)\\
& \propto &\exp \left\{-\frac{1}{2\sigma_r^2} \sum_{t=1}^T \sum_{i=1}^S(y_{itr}-\BM{x}_{it}^{\prime}\BM{b}_{r.})^{\prime} (y_{itr}-\BM{x}_{it}^{\prime}\BM{b}_{r.})\right\} \exp \left(-\frac{1}{2} \BM{b}_{r.}^{*^{\prime}} \BM{\Upsilon}_{1:(r-1),1:(r-1)}^{-1} \BM{b}_{r.}^*\right)\\
& \propto &\exp \left\{-\frac{1}{2\sigma_r^2} \sum_{t=1}^T \sum_{i=1}^S(y_{itr}-x_{itr}-\BM{x}_{it,1:(r-1)}^{\prime}\BM{b}_{r.}^*)^{\prime} (y_{itr}-x_{itr}-\BM{x}_{it,1:(r-1)}^{\prime}\BM{b}_{r.}^*)\right\} \\
& &\exp \left(-\frac{1}{2} \BM{b}_{r.}^{*^{\prime}} \BM{\Upsilon}_{1:(r-1),1: (r-1)}^{-1} \BM{b}_{r.}^*\right)\\
& \propto &\exp\left[-\frac{1}{2}\left\{\BM{b}_{r.}^{*^{\prime}}\left(\BM{\Upsilon}_{1:(r-1),1:(r-1)}^{-1} + \frac{1}{\sigma_r^2}\sum_{t=1}^T \sum_{i=1}^S\BM{x}_{it, 1:(r-1)} \BM{x}_{it, 1:(r-1)}^{\prime}\right) \BM{b}_{r.}^* \right.\right.\\
& & \left. \left. \ \ \ \ \ \ \ \ \ \ \ \ \ \ \ -2 \frac{1}{\sigma_r^2} \BM{b}_{r.}^{*^{\prime}} \sum_{t=1}^T \sum_{i=1}^S\BM{x}_{it, 1:(r-1)}(y_{itr}-x_{itr})\right\}\right].
\end{eqnarray*}
% \end{autobreak}
% \end{eqnarray*}
% $}
% \end{equation*}

Therefore, when $1<r<L$, the full conditional distribution of $\BM{b}_{r.}^*$ is $N(\mb{m}_r^b,\mb{C}_r^b)$, where 
\begin{eqnarray*}
\mb{C}_r^b & = & \left(\BM{\Upsilon}_{1:(r-1),1:(r-1)}^{-1} + \sigma_r^{-2}\sum_{t=1}^T \sum_{i=1}^S\BM{x}_{it, 1:(r-1)} \BM{x}_{it, 1:(r-1)}^{\prime}\right)^{-1}\\
\mb{m}_r^b & = & \mb{C}_r^b \sigma_r^{-2} \sum_{t=1}^T \sum_{i=1}^S\BM{x}_{it, 1:(r-1)}(y_{itr}-x_{itr}).  
\end{eqnarray*}

When $r > L$, the full conditional density of $\BM{b}_{r.}$ is
% \begin{equation*}
% \resizebox{.95\columnwidth}{!}{$
\begin{eqnarray*}
p(\mb{b}_{r.} \mid-) & \propto&  \exp \left\{-\frac{1}{2} \sum_{t=1}^T \sum_{i=1}^S(\BM{y}_{it}-\BM{B}\BM{x}_{it})^{\prime} \BM{V}^{-1}(\BM{y}_{it}-\BM{B} \BM{x}_{it})\right\} p(\BM{b}_{r.})\\
& \propto &\exp \left\{-\frac{1}{2\sigma_r^2} \sum_{t=1}^T \sum_{i=1}^S(y_{itr}-\BM{x}_{it}^{\prime}\BM{b}_{r.})^{\prime} (y_{itr}-\BM{x}_{it}^{\prime}\BM{b}_{r.})\right\} \exp \left(-\frac{1}{2} \BM{b}_{r.}^{\prime} \BM{\Upsilon}^{-1} \BM{b}_{r.}\right)\\
% & \propto&  \exp (-\frac{1}{2} \sum_{t=1}^T\sum_{i=1}^S(\mb{y}_{it}-\mb{B} \mb{x}_{it})^{\prime} \mb{V}^{-1}(\mb{y}_{it}-\mb{B} \mb{x}_{it})) \exp (-\frac{1}{2} \mb{b}_{r.}^{\prime} \BM{\Upsilon}^{-1} \mb{b}_{r.}) \\
% & \propto&  \exp (-\frac{1}{2 \sigma_r^2}(\mb{y}_{., r}-\mb{X} \mb{b}_{r.})^{\prime}(\mb{y}_{., r}-\mb{X} \mb{b}_{r.})) \exp (-\frac{1}{2} \mb{b}_{r.}^{\prime} \BM{\Upsilon}^{-1} \mb{b}_{r.}) \\
%& \propto&  \exp \left[-\frac{1}{2}\left\{\mb{b}_{r.}^{\prime} \BM{\Upsilon}^{-1} \mb{b}_{r.}+\left(\frac{1}{\sigma_r^2}\sum_{t=1}^T \sum_{i=1}^S(y_{itr}^{\prime}-\mb{b}_{r.}^{\prime} \mb{x}_{it})(y_{itr}-\mb{x}_{it}^{\prime}\mb{b}_{r.}\right)\right\}\right] \\
& \propto&  \exp\left[-\frac{1}{2}\left\{\mb{b}_{r.}^{\prime}\left(\BM{\Upsilon}^{-1}+\frac{1}{\sigma_r^2}\sum_{t=1}^T \sum_{i=1}^S \mb{x}_{it} \mb{x}_{it}^{\prime}\right) \mb{b}_{r.}-\frac{2}{\sigma_r^2} \mb{b}_{r.}^{\prime}\sum_{t=1}^T \sum_{i=1}^S \mb{x}_{it} y_{itr}\right\}\right].
\end{eqnarray*}
% $}
% \end{equation*}
Therefore, when $r > L$,  the full conditional distribution of $\mb{b}_{r.}$ is $N(\BM{m}_r^b, \BM{C}_r^b )$, where
\begin{eqnarray*}
\mb{C}_r^b & = & \left(\BM{\Upsilon}^{-1}+\frac{1}{\sigma_r^2}\sum_{t=1}^T \sum_{i=1}^S \mb{x}_{it} \mb{x}_{it}^{\prime}\right)^{-1}\\
\mb{m}_r^b & = & \mb{C}_r^b \sigma_r^{-2} \sum_{t=1}^T \sum_{i=1}^S \mb{x}_{it} y_{itr}.  
\end{eqnarray*}

\paragraph{Uniquenesses $\sigma_r^2$}
%As $\BM{V}$ is the covariance matrix of the idiosyncratic error of the Bayesian factor models term, we assume that each element is independent of the others. As a variance, its prior follows an inverse gamma distribution. We can take a look at the $j$-th diagonal value first. Since the elements are independent of each other, we can use the corresponding columns of $\BM{Y}$ and $\BM{XB}^T$ to obtain the likelihood. The $\BM{Y}$ and $\BM{X}$ here are $n\times R$ and $n \times L$ matrices, where $n = S\times T$. 
The full conditional density of $\sigma_r^2$ is
\begin{eqnarray*}
% \resizebox{.95\columnwidth}{!}{$
p(\sigma_r^2| \cdot) \propto (\sigma_r^2)^{-(ST+n_\sigma)/2 -1} \exp \left[ -\frac{1}{2 \sigma_r^2}\left\{n_\sigma s_\sigma^2 + \sum_{t=1}^T\sum_{i=1}^S(y_{itr}-\BM{x}_{it}^\prime\BM{b}_{r.})'(y_{itr}-\BM{x}_{it}^\prime\BM{b}_{r.})\right\}\right].    
% $}.
\end{eqnarray*}
Therefore, the full conditional distribution of $\sigma_r^2$ is $IG(0.5\{ST+n_\sigma\},0.5\{n_\sigma s_\sigma^2 + \sum_{t=1}^T\sum_{i=1}^S(y_{itr}-\BM{x}_{it}^\prime\BM{b}_{r.})^2\})$.

\paragraph{Factor loadings variance $\tau_l$}
For $l=1,\dots,L$, let $\BM{B}_{(l+1):R,l}$ be the $l$-th column of $\BM{B}$ with the first $l$ elements removed. 
Then, the full conditional density of $\tau_l$ is
% \begin{equation*}
% \resizebox{.95\columnwidth}{!}{$
\begin{eqnarray*}
p(.\tau_l|_{-}) & \propto&  p(\tau_l) p(\BM{B}_{(l+1):R,l} \mid \tau_l) \\
& \propto&  \frac{n_\tau s_\tau^2 / 2}{\Gamma(n_\tau / 2)} \tau_l^{-(n_\tau / 2+1)} \exp \left(-\frac{n_\tau s_\tau^2}{2\tau_l}\right) \tau_l^{-(R-i) / 2} \exp \left(-\frac{1}{2 \tau_l} \BM{B}_{(l+1):R,l}^{\prime} \BM{B}_{(l+1):R,l}\right) \\
& \propto&  \tau_l^{-(R-l+n_\tau) / 2-1} \exp \left[-\frac{1}{2 \tau_l}\left(\BM{B}_{(l+1):R,l}^{\prime} \BM{B}_{(l+1):R,l}+n_\tau s_\tau^2\right)\right].
\end{eqnarray*}
% $}
% \end{equation*}
Therefore, the full conditional distribution of $\tau_l$ is $IG ((R-l+n_\tau) / 2,(\BM{B}_{(l+1):R,l}^{\prime} \BM{B}_{(l+1):R,l}+n_\tau s_\tau^2) / 2)$.

\paragraph{Initial cluster probability $p_g$}
At time $t=1$, the probability that subject $i$ belongs to cluster $g$ is $p_g = P(Z_{i1} = g)$. 
Then, the full conditional density of $\BM{p}=(p_1,\dots,p_G)$ is
\begin{eqnarray*}
    p(p_1,\dots,p_G|.) & \propto&  p_1^{\alpha_1 - 1}\times \dots\times p_G^{\alpha_G - 1}\times\prod_{i=1}^S p_{Z_{i1}} = \prod_{g=1}^G p_{g}^{n_{1g}+\alpha_g-1}.
\end{eqnarray*}
Therefore, the full conditional distribution of $(p_1,\dots,p_G)$ is $Dirichlet(\alpha_1^*,\dots,\alpha_G^*)$ where $\alpha_{g}^* = n_{1g}+\alpha_g$, $g=1,\dots,G$.

\paragraph{Transition probability matrix $q_{j1},\dots,q_{jG}$}
The full conditional density of $(q_{j1},\dots,q_{jG})$ is
\begin{eqnarray*}
p(q_{j1},\dots,q_{jG}|.) &\propto&  q_{j1}^{\alpha_{j1}-1}\times\dots \times q_{jG}^{\alpha_{jG}-1}\times\prod_{i=1}^{S} \prod_{t=2}^{T} \prod_{g=1}^G ( P(Z_{it}=g|Z_{i,t-1}=j) )^{\mathds{1}(Z_{it}=g \mbox{ and } Z_{i,t-1}=j)}\\
& \propto&  \prod_{g=1}^{G} q_{jg}^{\sum_{i=1}^{S} \sum_{t=2}^{T} \mathds{1}(Z_{it}=g \mbox{ and } Z_{i,t-1}=j)+\alpha_{jg}-1}\\
& =&  \prod_{g=1}^G q_{jg}^{m_{jg}+\alpha_{jg}-1},
\end{eqnarray*}
where $m_{jg} = \sum_{i=1}^S \sum_{t=2}^T \mathds{1}(Z_{it}=g \mbox{ and } Z_{i,t-1}=j)$. 
Therefore, the full conditional distribution of $(q_{j1},\dots,q_{jG})$ is $Dirichlet(m_{j1}+\alpha_{j1},\dots,m_{jG}+\alpha_{jG})$.

\paragraph{Cluster assignment $Z_{it}$}
Conditional on $Z_{i,t-1}=j$ and $Z_{i,t+1}=p$, the full conditional density of $Z_{it}$ is
% \begin{equation*}
% \resizebox{.95\columnwidth}{!}{$
\begin{eqnarray*}
    p(Z_{it}=g|.) &\propto&  f(\BM{x}_{it}|Z_{it}=g) P(Z_{it}=g|Z_{i,t-1}=j)P(Z_{i,t+1}=p|Z_{it}=g)\\
    & \propto&  |\BM{\Omega}_g|^{-1/2}\exp\left[-\frac{1}{2}(\BM{x}_{it}-\BM{\mu}_g)'\BM{\Omega}_g^{-1}(\BM{x}_{it}-\BM{\mu}_g)\right] q_{jg} q_{gp}, 
\end{eqnarray*}
% $}
% \end{equation*}
where $g=1,\dots,G$. Thus, this is a discrete distribution with constant of proportionality equal to $\{\Sigma_{g=1}^G|\BM{\Omega}_g|^{-1/2}\exp[-0.5(\BM{x}_{it}-\BM{\mu}_g)'\BM{\Omega}_g^{-1}(\BM{x}_{it}-\BM{\mu}_g)]q_{jg}q_{gp}\}^{-1}$.

% \begin{figure}[H]
%     \centering
%     \includegraphics[width=\linewidth]{proj1/sim-data/sigma2_CI_g4l3s200times5.pdf}
%     \caption{Simulated data -- uniqueness variances for the 4-cluster-3-factor using BDCFM: 95\% credible interval (vertical line) and posterior mean (black dot), and true values (red dashed line).}
%     \label{fig:simulatedsigma}
% \end{figure}

%\label{lastpage}

\end{document}